\documentclass[english]{eptcs}
\providecommand{\event}{SAIRP 2013}
\usepackage[latin1]{inputenc}
\usepackage[T1]{fontenc}
\usepackage{babel}
\usepackage{graphicx}
\usepackage{amsmath}
\usepackage{amsfonts}
\usepackage{amssymb}
\usepackage{multirow}

\usepackage{url}
\usepackage{stmaryrd}

\newcommand{\bbL}{\{\hspace{-3pt}\{}
\newcommand{\bbR}{\}\hspace{-3pt}\}}

\newcommand{\aaL}{\langle{\hspace*{-2.5pt}}\langle}
\newcommand{\aaR}{\rangle{\hspace*{-2.5pt}}\rangle}

\newcommand{\andw}
{\,{\scriptstyle{\wedge}}\hspace{-5.2pt}{\scriptstyle{\wedge}}\,} 
\newcommand{\orv}
{\,{\scriptstyle{\vee}}\hspace{-5.2pt}{\scriptstyle{\vee}}\,} 

\usepackage{color}

\newtheorem{theorem}{Theorem}

\title{Verification of Imperative Programs\\
by Constraint Logic Program Transformation}

\author{
Emanuele De Angelis
\institute{DEC, University `G. D'Annunzio', Pescara, Italy}
\email{emanuele.deangelis@unich.it}
\and
Fabio Fioravanti
\institute{DEC, University `G. D'Annunzio', Pescara, Italy}
\email{fioravanti@unich.it}
\and 
Alberto Pettorossi\hspace{6mm}
\institute{DICII, University of Rome Tor Vergata, Rome, Italy}\hspace{5mm}
\email{pettorossi@disp.uniroma2.it}\hspace{4mm}
\and
Maurizio Proietti\hspace{9mm}
\institute{IASI-CNR, Rome, Italy}\hspace{9mm}
\email{maurizio.proietti@iasi.cnr.it}\hspace{10mm}
}

\def\titlerunning{Verification of Imperative Programs by Constraint Logic Program Transformation}
\def\authorrunning{E.~De~Angelis, F.~Fioravanti, A.~Pettorossi, and M.~Proietti}

\begin{document}

\maketitle

\begin{abstract}
We present a method for verifying partial correctness properties of
imperative programs that manipulate integers and arrays by using techniques based on the
transformation of constraint logic programs~(CLP).
We use CLP
as a metalanguage for representing imperative programs, their executions, and
their properties.
First, we encode the correctness of an imperative program, say \textit{prog},
as the negation of a predicate $\mathtt{incorrect}$ defined
by a CLP program~$T$. 
By construction, $\mathtt{incorrect}$ holds in
the least model of~$T$ if and only if 
the execution of \textit{prog} 
from an initial configuration 
eventually halts in
an error configuration.
Then, we apply to program~$T$ a sequence of  transformations 
that preserve its least model semantics. These transformations are based on
well-known transformation {\em rules}, such as {\em unfolding} and {\em folding},
guided by suitable transformation {\em strategies}, such as {\em specialization}
and {\em generalization}.
The objective of the transformations is to derive a new CLP program~$\mathit{TransfT}$
where the predicate {\tt incorrect} is defined either by (i) the fact
`{\tt incorrect.}' (and in this case \textit{prog} is {\em not} correct), 
or by (ii) the empty set of clauses (and in this case \textit{prog} is correct).
In the case where we derive a CLP program such that neither (i) nor (ii) holds, 
we iterate the transformation.
Since the problem is undecidable, this process may not terminate.
We show through examples that our method can be applied
in a rather systematic way, and is amenable to automation
by transferring to the field of program verification many techniques
developed in the field of program transformation.
\end{abstract}

\pagestyle{plain} 

\section{Introduction}
\label{sec:intro}
In the last decade formal techniques have received a renewed attention as the 
basis of a methodology for increasing the reliability of software artifacts and reducing the cost of software production~\cite{Mi&10}. 
In particular, a massive effort has been made to devise automatic verification 
techniques, such as
{\em software model checking}~\cite{JhM09}, 
for proving the correctness of programs with respect to
their specifications.  

In many software model checking techniques, 
the notion of a {\em constraint} has been shown to be very
effective, both for constructing models of programs and
for reasoning about them~\cite{Be&07,CoH78,De&13a,DeP99,Fla04,Gr&12,Ja&11b,Pe&98,PoR07}.
Several types of constraints have been considered, such as 
equalities and inequalities over the booleans, the integers, the reals, 
the finite or infinite trees.
By using constraints we can represent 
in a symbolic, compact way
(possibly infinite) sets of values
computed by programs and, more in general, sets of states
reached during program executions.
Then, in order to reason about program properties in an efficient way,
we can use solvers specifically designed for the various classes of
constraints we have mentioned above. 

In this paper we consider a simple imperative programming language
with integer and array variables, and we
adopt Constraint Logic Programming (CLP)~\cite{JaM94}
as a metalanguage for representing imperative programs, their executions, and
their properties.
We use constraints consisting of equalities and inequalities over
the integers, but the method presented here is parametric with respect to the 
constraint domain which is used.
By following an approach first presented in~\cite{Pe&98},
a given imperative program~\textit{prog}
and its interpreter 
are encoded as a CLP program.
Then, the proofs of the properties of interest about the program~\textit{prog}
are sought by analyzing the derived CLP program.
In order to improve the efficiency of analysis, it is advisable to
first {\em compile-away} the CLP interpreter of the language in  which~\textit{prog}
is written. This is done by specializing the interpreter with respect to the given program~\textit{prog} using well-known {\em program specialization} 
(also known as {\em partial evaluation}) techniques~\cite{Jo&93,Pe&98}.
 
We have shown in previous work~\cite{De&13b,De&13a,Fi&11b}
that program specialization can be
used not only as a preprocessing step to improve the efficiency
of program analysis, but also as a means of analysis on its own.
In this paper, we extend that approach and we propose a verification
methodology based on more general, semantics preserving 
{\em unfold/fold transformation rules} for CLP programs~\cite{BuD77,EtG96,TaS84}.

Transformation-based verification techniques are very appealing
because they are parametric with respect to both the programming languages 
in which programs are written, and the logics in which the properties of interest
are specified.
Moreover, since the output of a transformation-based verification of a program
is an \textit{equivalent} program with respect to the properties of 
interest, we can apply a \textit{sequence} of transformations, thereby refining 
the analysis to the desired degree of precision.

Our approach can be summarized as follows.
Suppose we are given: (i) an imperative program~\textit{prog},
(ii)~a predicate \texttt{initConf} expressing the property, called the {\em initial property}, 
holding in the configuration from which the execution of~\textit{prog} starts, and
(iii)~a predicate \texttt{errorConf} expressing the property, called the {\em error property}, 
holding in the configuration which should {\em not} be reached at the end of the execution 
of~\textit{prog}. The partial correctness of \textit{prog}  
is defined as the negation of a predicate 
$\mathtt{incorrect}$ specified by the following CLP program $T$:\nopagebreak

\smallskip

$\mathtt{incorrect}$ {\tt{:-}} $\mathtt{initConf(X) ,\   reach(X).}$\nopagebreak

$\mathtt{reach(X)}$ {\tt{:-}} $\mathtt{tr(X,X1) ,\  reach(X1).}$\nopagebreak

$\mathtt{reach(X)}$ {\tt{:-}} $\mathtt{errorConf(X).}$

\smallskip
\noindent
where: (i)~$\texttt{reach(X)}$ holds iff an error configuration can be
reached from the configuration $\texttt{X}$,
and (ii)~$\mathtt{tr(X,X1)}$ holds iff the configuration $\texttt{X1}$ 
can be reached in one step from the configuration $\texttt{X}$ via
the transition relation that defines the operational semantics of the
imperative language.
Thus, $\mathtt{incorrect}$ holds iff there exists an error configuration that
can be reached from an initial configuration.

The verification method we propose in this paper is made out of 
 the following two steps. 

\noindent {\it Step}~(A). This step, called the {\em removal of the interpreter}, consists in
specializing the program $T$ (which includes the clauses defining the
predicate $\mathtt{tr}$)
with respect to the given program \textit{prog} and properties
\texttt{initConf} and \texttt{errorConf}. After this specialization step
we derive from program~$T$ a new program~$T_A$,
where there is no reference to the predicate~$\mathtt{tr}$ 
(and in this sense we say that during this Step~(A) the interpreter is removed or
`compiled-away').

\noindent {\it Step}~(B). This step, called the {\em propagation of the initial and error properties},
 consists in applying a sequence 
of unfold/fold transformation rules, and deriving from the CLP 
program~$T_A$ a new CLP program $T_B$ such that ${\tt incorrect}$ holds in $T_B$
if and only if \textit{prog} is {\em not} 
correct with respect to the given initial and error
properties.
The objective of this Step~(B) is to derive a program $T_B$ 
where the predicate {\tt incorrect} is defined by: either (i)~the fact
`{\tt incorrect.}' (and in this case \textit{prog} is not correct), 
or (ii)~the empty set of clauses (and in this case \textit{prog} is correct).
If neither Case~(i) nor Case~(ii) holds, that is,  
in program $T_B$ the predicate {\tt incorrect} is defined by a non-empty set of clauses
which does not contain the fact `{\tt incorrect.}', we can conclude 
neither the correctness nor the incorrectness of~\textit{prog}. Thus, similarly to what 
has been proposed in~\cite{De&13a}, we continue our verification method by
iterating this Step~(B)
in the hope of deriving a program where either Case~(i) or Case~(ii) holds. 
Obviously, due to undecidability limitations, 
it may be the case that we never derive such a program.

\smallskip
During Step (B) we apply transformation rules that are more powerful than those
needed for program specialization and, in particular, for the removal of the interpreter
done during Step~(A).
Indeed, the rules used during Step~(B) include the
{\em conjunctive definition} and the {\em conjunctive folding} rules and they
 allow us to introduce and transform 
new predicate definitions that correspond
to \textit{conjunctions} of old predicates, while program specialization 
can deal only with new predicates that correspond to specialized versions
of \textit{one} old predicate. During Step~(B) we use also
the {\em goal replacement} rule,
which allows us to replace conjunctions of constraints and predicates 
by applying equivalences that hold in the least model of $T$,
while program specialization can only replace conjunctions of constraints.

These more powerful rules extend the specialization-based
verification techniques in two ways. 
First, we can verify programs with respect to complex initial and error properties defined by sets of CLP clauses 
(for instance, recursively defined relations among
program variables), whereas program specialization
can only deal with initial and error properties 
specified by (sets of) constraints.
Second, we can verify programs which manipulate arrays and other
data structures by applying equivalences between predicates 
that express suitable properties of those structures.
In particular, in this paper we will apply equivalences which follow from the 
axiomatization of the theory of arrays~\cite{Br&06}. 

\smallskip

The paper is organized as follows.
In Section~\ref{sec:syntax_semantics} we present the imperative 
language and the definition of its interpreter as a CLP program.
In Section~\ref{sec:translating-correctness-into-CLP} we describe how
partial correctness properties of imperative programs can be translated 
to predicates defined by CLP programs.
In Section~\ref{sec:verification_method} we present our transformation-based
verification method, and a general strategy to apply the transformation rules.
Next, we present two examples of application of our verification
method. In particular, in Section~\ref{sec:conjunctivefolding} we show how we 
deal with specifications provided by 
recursive CLP clauses, and in Section~\ref{sec:array_properties} we show how we 
deal with programs that manipulate arrays.
Finally, in Section~\ref{sec:conclusions} we discuss related work in the
area of program verification.


\section{Translating Imperative Programs into CLP}
\label{sec:syntax_semantics}

We consider a simple imperative language with arrays. We are given: (i)~the set {\it{Vars}} of
{\rm{integer variable identifiers}}, 
(ii)~the set~{\it{AVars}}  of {\rm{integer array identifiers}}, 
and (iii)~the set~$\mathbb{Z}$  of the integer constants.
Our language has the following syntax:

\vspace{1mm}
\hspace{-7mm}
{\rm{
\begin{tabular}{l@{\hspace{0.5mm}}c@{\hspace{1mm}}lllll}
$x,y,z,i,j,m,n,\ldots~ $ & $\in$ & $\mathit{Vars}$ (integer variable identifiers) & \\
$a,b,\ldots$& $\in$ & $\mathit{AVars}$ (integer array identifiers) & \\
$\mathit{k,\ldots}$ & $\in$ & $\mathbb{Z}$ (integer constants)    & \\
$\mathit{\ell}, \mathit{\ell}_{0}, \mathit{\ell}_{1}, \ldots $ & $\in$ & Labels ($\subseteq \mathbb{Z}$) & \\
$\mathit{uop}$, $\mathit{bop}$ & $\in$ & Unary and binary operators ({\tt{-}}, 
{\tt{+}}, {\tt{<}}, $\ldots$) & \\[1.5mm]
\end{tabular}
}}

\hspace{-7mm}
{\rm{
\begin{tabular}{l@{\hspace{1.5mm}}c@{\hspace{2mm}}lllll}
$ {\mathrm{prog}}$  & ::= & 
$ {\mathrm{lab\!\_\!cmd}} ^{+}$ & &\\
\makebox[13mm][l]{$ {\mathrm{lab\!\_\!cmd}}$} & ::= & 
$\ell\! :\!  {\mathrm{cmd}} $ \\
\end{tabular}
}}

\vspace{-0.5mm}
\hspace{-7mm}
{\rm{
\begin{tabular}{l@{\hspace{1.5mm}}c@{\hspace{2mm}}lll}
\makebox[13mm][l]{${\mathrm{cmd}}$}& ::= & 
${\mathtt{halt}} ~~|~~ 
 x\! =\!  {\mathrm{expr}}  ~~|~~ a${\tt[}$\mathrm{expr}${\tt]}$=\!\mathrm{expr}$ 
  ~~$|$~~  ${\mathtt{if~}}${\tt(}$ {\mathrm{expr}}${\tt)}$~\ell_{1}~{\mathtt{else}}$ 
$\ell_{2}$ ~~$|$~~ ${\mathtt{goto}}$ $\ell$  
\end{tabular}
}}

\vspace{-0.5mm}
\hspace{-7mm}
{\rm{
\begin{tabular}{l@{\hspace{1.5mm}}c@{\hspace{2mm}}ll}
\makebox[13mm][l]{${\mathrm{expr}}$}& ::= & $\emph{k}  ~~|~~ \emph{x}  ~~|~~ \emph{uop}
 ~ {\mathrm{expr}}  ~~|~~ {\mathrm{expr}} $ \emph{bop} $ {\mathrm {expr}}   ~~|~~ 
a${\tt[}$\mathrm{expr}${\tt]}$ $ \\
\end{tabular}
}}

\smallskip

\noindent
A program is a non-empty sequence of labeled commands (also called
commands, for brevity). 
The elements of a sequence are separated by semicolons. 
We assume that in every program each label occurs at most once. 
Note that in our language we can deal with conditional and iterative commands, such as
`${\mathtt{if~}}${\tt(}$ {\mathrm{expr}}${\tt)}~{cmd}', 
`${\mathtt{if~}}${\tt(}$ {\mathrm{expr}}${\tt)}~{cmd}~${\mathtt{else}}$ {cmd}',
and `${\mathtt{while~}}${\tt(}$ {\mathrm{expr}}${\tt)}~\{{cmd}\}', 
 by translating them using
${\mathtt{if}}$-${\mathtt{else}}$ and ${\mathtt{goto}}$ commands.

In order to focus our attention on the verification issues,
we do not consider in our imperative language other features such as:
 (i)~type and variable declarations, (ii)~function
declarations and function calls, 
and (iii) multidimensional arrays.
Those features can be added in a straightforward way (see, for instance,~\cite{De&13a}). 

\smallskip
Now we give the semantics of our language
by defining a binary relation $\Longrightarrow$
which will be encoded by a CLP predicate {\tt tr}.
For that purpose let us first
introduce the following auxiliary notions.

An {\it{environment}} $\delta$ is a function 
that maps: (i)~every integer variable identifier $x$ to its value $v\in \mathbb Z$,
and (ii)~every array identifier $a$ to a finite function from 
the set $\{0,\ldots,{\textit{dim}}(a)\!-\!1\}$ to $\mathbb Z$, 
where ${\mathit{dim}}(a)$ is the dimension of $a$.
For any expression~$\mathit{e}$, 
array identifier $a$,
and environment~$\delta$, 
(i) $\llbracket \mathit{e}\rrbracket\delta$ is the integer value of $\mathit{e}$
defined by induction on the structure of $\mathit{e}$ (in particular, for any integer
variable identifier~$x$,
$\llbracket x\rrbracket\delta =_\textit{def} \delta(x)$),
and (ii) $\llbracket a[\mathit{e}]\rrbracket\delta=_\textit{def}\delta(a)(\llbracket \mathit{e}\rrbracket\delta)$.
We assume that the evaluation of expressions has no side effects.

A {\it{configuration}} is a pair of the form
$\aaL c, \delta\aaR$ where: (i)~$c$~is a labeled command, and
(ii)~$\delta$ is an environment.
By $\mathit{update}(f,x,v)$ we denote the function $f'$ such that, for every $y$,
if $y\!=\!x$ then $f'\!(y)\!=\!v$ else $f'\!(y)\!=\!f(y)$.
By $\mathit{write}(f,x,v)$ we denote the function $\mathit{update}(f,x,v)$
in the case where $f$ is a finite function 
denoting an array, $x$ is an integer in the domain of $f$, and $v$ is an integer value. 
For any program~{\textit{prog}}, for any label $\ell$, (i)~${\mathit{at}}(\ell)$ denotes
the command in~{\textit{prog}} with label~$\ell$, and
(ii)~{\it{nextlab}}$(\ell)$ denotes
the label of the command in~{\textit{prog}} which is written 
{\it{immediately after}} the command 
with label~$\ell$.

\smallskip
The operational semantics (that is, the interpreter) 
of our language is given as a transition
relation $\Longrightarrow$ between
configurations according to the 
following rules $R1$--$R3$. 
Notice that no rules are given for the command
$\ell\!:\! {\mathtt{halt}}$. 
Thus, no configuration $\gamma$ exists such that 
$\ell\!:\! \mathtt{halt}\, \Longrightarrow\, \gamma$.

\smallskip
\noindent
$(R1)$. {\it{Assignment}}. 

\smallskip
\noindent
If \emph{$x$} is an integer variable identifier:

\noindent
\makebox[6mm][l]{}$\aaL  \ell\!:\!x\!=\!e,\ \delta\aaR \Longrightarrow
\aaL{\mathit{at}}(\!{\mathit{nextlab}}(\ell)\!),\ {\mathit{update}}(\delta,x,\llbracket e\rrbracket\delta)\aaR$

\smallskip
\noindent
If \emph{$a$} is an integer array identifier:

\noindent
\makebox[6mm][l]{}$\aaL  \ell\!:\!a[\mathit{ie}]\!=\!e,\ \delta\aaR \Longrightarrow
\aaL{\mathit{at}}(\!{\mathit{nextlab}}(\ell)\!),\ 
{\mathit{update}}(\delta, a,\mathit{write}(\delta(a),\llbracket \mathit{ie}\rrbracket\delta,\llbracket e\rrbracket\delta))\aaR$

\smallskip
\noindent

\smallskip
\noindent
\makebox[22mm][l]{$(R2)$. {\it{Goto}}.}
$\aaL \ell\!:\!{\mathit{goto}}~\ell',\ \delta\aaR\Longrightarrow\aaL
{\mathit{at}}(\ell'),\ \delta\aaR$

\smallskip
\noindent
$(R3)$. {\it{If-else}}.

\smallskip
\noindent
\makebox[22mm][l]{If $\llbracket e\rrbracket\delta\!\not =\!0$:} 
$\aaL \ell\!: {\mathtt{if}}~${\tt(}$e${\tt)}$~ \ell_{1}~{\mathtt{else}}~\ell_{2},
\ \delta\aaR
\Longrightarrow\aaL{\mathit{at}}(\ell_{1}),\ \delta\aaR$

\smallskip
\noindent
\makebox[22mm][l]{If $\llbracket e\rrbracket\delta\!=\!0$:} 
$\aaL \ell\!: {\mathtt{if}}~${\tt(}$e${\tt)}$~ \ell_{1}~{\mathtt{else}}~\ell_{2},
\ \delta\aaR
\Longrightarrow\aaL{\mathit{at}}(\ell_{2}),\ \delta\aaR$

\smallskip 

Let us now recall some notions and terminology concerning constraint 
logic programming (CLP). For more details the reader may refer to~\cite{JaM94}.
If $\mathtt{p_1}$ and $\mathtt{p_2}$ 
are linear polynomials with integer coefficients and integer variables,
then $\mathtt{p_1\!=\!p_2}$, $\mathtt{p_1\!\not =\!p_2}$, $\mathtt{p_1\!<\! p_2}$, 
$\mathtt{p_1\!\leq\! p_2}$, $\mathtt{p_1\!\geq\! p_2}$, and 
$\mathtt{p_1\!>\!p_2}$ are {\em atomic constraints}.
A {\em constraint} is either $\tt{true}$, or $\tt{false}$,
or an atomic constraint, or a {\it{conjunction}} of constraints.
A CLP program is a finite set of clauses of the form 
$\texttt{A\,:-\,c,B}$,
where $\texttt{A}$ is an atom, $\texttt{c}$ is a constraint, 
and $\texttt{B}$ is a (possibly empty) conjunction of atoms.
A clause of the form: $\texttt{A\,:-\,c}$ is called a {\em constrained fact} or 
simply a {\em fact} if~$\texttt{c}$ is $\tt{true}$.

The semantics of a CLP program $P$ is defined to be the {\em least model} of $P$ 
which extends the standard interpretation on the integers.
This model is denoted by $M(P)$.

The CLP interpreter for our imperative
language is given by the following predicate $\mathtt{tr}$
which encodes the transition relation $\Longrightarrow$. 

\smallskip
\noindent    
\makebox[6mm][r]{1.~}~$\mathtt{tr(cf(cmd(L,asgn(int(X),E)),D),\ cf(cmd(L1,C),D1))}$ {\tt{:-}}\nopagebreak

~~~~$\mathtt{eval(E,D,V), ~~update(D,int(X),V,D1), ~~nextlab(L,L1), ~at(L1,C). }$

\noindent    
\makebox[6mm][r]{2.~}~$\mathtt{tr(cf(cmd(L,asgn(arrayelem(A,IE),E)),D),\ cf(cmd(L1,C),D1))}$ {\tt{:-}}\nopagebreak

~~~~$\mathtt{eval(IE,D,I), ~~eval(E,D,V), ~~lookup(D,array(A),FA), ~~write(FA,I,V,FA1),}$

~~~~$\mathtt{update(D,array(A),FA1,D1), ~~nextlab(L,L1),  ~~at(L1,C). }$

\noindent    
\makebox[6mm][r]{3.~}~$\mathtt{tr(cf(cmd(L,ite(E,L1,L2)),D),\ cf(cmd(L1,C),D))}$ {\tt{:-}} $\mathtt{V\neq0, ~~eval(E,D,V), ~~at(L1,C). }$
    
\noindent   
\makebox[6mm][r]{4.~}~$\mathtt{tr(cf(cmd(L,ite(E,L1,L2)),D),\ cf(cmd(L2,C),D))}$ {\tt{:-}} $\mathtt{V=0, ~~eval(E,D,V),  ~~at(L2,C). }$

\noindent       
\makebox[6mm][r]{5.~}~$\mathtt{tr(cf(cmd(L,goto(L1)),D),\ cf(cmd(L1,C),D))}$ {\tt{:-}} 
$\mathtt{at(L1,C).}$ 

\smallskip

\noindent
In the above clauses the term
$\mathtt{asgn(int(X),E)}$ encodes a variable assignment of the form
$x\!=\!e$, while $\mathtt{asgn(arrayelem(A,IE),E)}$ 
encodes an array assignment of the form
$a${\tt [}$\mathit{ie}${\tt ]}$=\!e$.
Similarly, the terms $\mathtt{ite(E,L1,L2)}$ and
$\mathtt{goto(L)}$ encode the conditional 
${\mathtt{if}}\,${\tt(}$e${\tt)}$\ \ell_{1}\ {\mathtt{else}}\ \ell_{2}$ 
and the jump ${\mathtt{goto}}\ \ell$, respectively.
The term $\mathtt{cmd(L,C)}$ encodes the command~$\mathtt{C}$  with label 
$\mathtt{L}$.
The predicate $\mathtt{at(L,C)}$ binds to $\mathtt{C}$ the command 
with label~$\mathtt{L}$.
The predicate $\mathtt{nextlab(L,L1)}$ binds to~$\mathtt{L1}$ the label of
the command 
which is written immediately after the command 
with label~$\mathtt{L}$. 

The predicate $\mathtt{eval(E,D,V)}$
computes the value 
$\mathtt{V}$ of the expression~$\mathtt{E}$ in the environment~$\mathtt{D}$.
Below we list a subset of the clauses that define  $\mathtt{eval(E,D,V)}$
by induction on the structure of~$\mathtt{E}$. The others are
similar and we shall omit them.

\smallskip

\noindent
\makebox[6mm][r]{6.~}~$\mathtt{eval(int(X),D,V)}$ {\tt{:-}}  $\mathtt{lookup(D,int(X),V).}$ 

\noindent
\makebox[6mm][r]{7.~}~$\mathtt{eval(not(E),D,V)}$ {\tt{:-}}  $\mathtt{{V\!\not =\!0}, ~~V1\!=\!0, ~~eval(E,D,V1).}$ 

\noindent
\makebox[6mm][r]{8.~}~$\mathtt{eval(not(E),D,V)}$ {\tt{:-}}  $\mathtt{V\!=\!0, ~~V1\!\neq\!0, ~~eval(E,D,V1).}$ 

\noindent
\makebox[6mm][r]{9.~}~$\mathtt{eval(plus(E1,E2),D,V)}$ {\tt{:-}}  $\mathtt{V=V1\!+\!V2, ~~eval(E1,D,V1), ~~eval(E2,D,V2).}$ 

\noindent
\makebox[6mm][r]{10.~}~$\mathtt{eval(arrayelem(A,IE),D,V)}$ {\tt{:-}}  
$\mathtt{eval(IE,D,I), ~~lookup(D,array(A),FA), ~~read(FA,I,V).}$

\smallskip 

\noindent
The predicate $\mathtt{update(D,Id,B,D1)}$ updates 
the environment~$\mathtt{D}$, thereby constructing 
the new  environment~$\mathtt{D1}$
by binding the (integer or array) identifier~$\mathtt{Id}$ to the (integer or {array}) 
value~$\mathtt{B}$. 
The predicate $\mathtt{lookup(D,Id,B)}$ retrieves from the environment $\mathtt{D}$ the (integer or {array}) value 
$\mathtt{B}$ bound to the identifier~$\mathtt{Id}$.
 
The predicate 
{\tt read} gets the value of an element of an
array, and the predicate 
{\tt write} sets the value of an element of an array.
As already mentioned, the environment maps an array to a finite function.
We represent this function as a pair $\texttt{(A,N)}$, where 
$\texttt{N}$ is the dimension of the array and
$\texttt{A}$ is a list of integers of length $\texttt{N}$.
The predicate ${\texttt{read((A,N),I,X)}}$ holds iff 
the~{\tt I}-th element of~{\tt A} is~{\tt X},
and the predicate ${\texttt{write((A,N),I,X,(A1,N))}}$ 
holds iff the list~{\tt A1} is equal
to~{\tt A} except that the {\tt I}-th element of~{\tt A1} 
is~{\tt X}.

For reasons of space, we do not list the clauses defining $\mathtt{update}$,
$\mathtt{lookup}$, $\mathtt{read}$, and $\mathtt{write}$. 

\section{Translating Partial Correctness into CLP}
\label{sec:translating-correctness-into-CLP}

The problem of verifying the partial
{\rm correctness} of a {\rm{program}} {\textit{prog}} is the problem of checking 
whether or not, starting from an initial
configuration, the execution of {\textit{prog}} leads to
an {\rm{error configuration}}. 
This problem is formalized by defining an {\it{incorrectness triple}} of the form
$\bbL\varphi_{\mathit{init}}\bbR $ {\textit{prog}} $
\bbL\varphi_{\mathit{error}}\bbR$,
where:

\hangindent=7mm
\noindent
\makebox[7mm][l]{(i)~}{\textit{prog}} is a program which acts on the variables
$z_1,\ldots,z_r$, each of which is either an integer variable or an 
integer array,

\hangindent=7mm
\noindent
\makebox[7mm][l]{(ii)~}$\varphi_{\mathit{init}}$ is a 
first-order formula
with free variables in $\{z_1,\ldots,z_r\}$
characterizing the initial configurations, and

\hangindent=7mm
\noindent
\makebox[7mm][l]{(iii)~}$\varphi_{\mathit{error}}$ is a 
first-order formula 
with free variables in $\{z_1,\ldots,z_r\}$
characterizing the error configurations.

\noindent
We assume that: (i)~the formulas $\varphi_{\mathit{init}}$ and $\varphi_{\mathit{error}}$
can be encoded using CLP predicates, and (ii)~the 
domain of every environment $\delta$ is the set $\{z_1,\ldots,z_r\}$. 
We also assume, without loss of generality, that the last command of 
{\textit{prog}} is $\ell_{h}\!:\!{\mathtt{halt}}$ and no other {\tt halt} command
occurs in {\textit{prog}}.

We say that a program  {\textit{prog}} is {\it{incorrect}} with respect to
a set of initial configurations 
satisfying $\varphi_{\mathit{init}}$
and a set of error configurations satisfying
$\varphi_{\mathit{error}}$, or simply,
${\textit{prog}}$ is {\it{incorrect}} with respect to
$\varphi_{\mathit{init}}$ and $\varphi_{\mathit{error}}$,
if there exist  environments
$\delta_{\mathit{init}}$ and $\delta_{\mathit{h}}$ such that

\smallskip
\noindent
\makebox[7mm][l]{(i)~}$\varphi_{\mathit{init}}(\delta_{\mathit{init}}(z_{1}),\ldots,\delta_{\mathit{init}}(z_{r}))$ holds,

\smallskip
\noindent
\makebox[7mm][l]{(ii)~}$\aaL \ell_{0}\!:\!c_{0}, 
\ \delta_{\mathit{init}}\aaR$ $\Longrightarrow ^*$
$\aaL \ell_{h}\!:\!{\mathtt{halt}},\  \delta_{\mathit{h}}\aaR$ and

\smallskip
\noindent
\makebox[7mm][l]{(iii)~}$\varphi_{\mathit{error}}(\delta_{\mathit{h}}(z_{1}),\ldots,\delta_{\mathit{h}}(z_{r}))$ holds,
 
\smallskip
\noindent
where $\ell_{0}\!:\!c_{0}$ is the first command in~{\textit{prog}}. Obviously, in 
$\varphi_{\mathit{init}}$ or in $\varphi_{\mathit{error}}$ some of the variables 
$z_{1},\ldots,z_{r}$  may be absent.
A program  is said to be {\it{correct}} with 
respect to $\varphi_{\mathit{init}}$ 
and $\varphi_{\mathit{error}}$ iff it is not incorrect
with respect to $\varphi_{\mathit{init}}$ 
and $\varphi_{\mathit{error}}$.

We now show, by means of an example, how to encode  
an incorrectness triple
as a CLP program.
The reader will have no difficulty to see how this
encoding can be done in general.
Let us consider the incorrectness triple 
$\bbL\varphi_{\mathit{init}}(i,j)\bbR$ {\it{increase}}
$\bbL\varphi_{\mathit{error}}(i,j)\bbR$
where:\nopagebreak

\noindent
--  $\varphi_{\mathit{init}}(i,j)$ is 
$ i\!=\!0 \andw j\!=\!0$,\nopagebreak

\noindent
\makebox[40mm][l]{-- {\it{increase}} is the program} 


\noindent
\hspace{-3mm}\makebox[6mm][l]{}$\ell_{0}$:~ $\mathtt{while}~(i\!<\!2n)~\{ 
\mathtt{if}~(i\!<\!n)~i=i\!+\!1;~\mathtt{else}~j=i\!+\!1;$\hspace{4mm}$i=i\!+\!1\}$\,;\nopagebreak

\noindent
\hspace{-3mm}\makebox[6mm][l]{}$\ell_{h}$:~ $\mathtt{halt}$\nopagebreak

\noindent
-- $\varphi_{\mathit{error}}(i,j)$ is  $ i\!<\!j$.

\smallskip
\noindent
First, we replace the given program {\textit{increase}} by the following sequence of commands:

\smallskip

\hspace{-3mm}$\ell_{0}$:~ $\mathtt{if}~(i\!<\!2n)~ \ell_{1} \mathtt{~else~} \ell_{h};$\nopagebreak

\hspace{-3mm}$\ell_{1}$:~ $\mathtt{if}~(i\!<\!n)~ \ell_{2} \mathtt{~else~} \ell_{4};$

\hspace{-3mm}$\ell_{2}$:~  $i=i\!+\!1;$\nopagebreak

\hspace{-3mm}$\ell_{3}$:~ $\mathtt{goto}~ \ell_{5};$\nopagebreak

\hspace{-3mm}$\ell_{4}$:~ $j=i\!+\!1;$\nopagebreak

\hspace{-3mm}$\ell_{5}$:~ $i=i\!+\!1;$\nopagebreak

\hspace{-3mm}$\ell_{6}$:~ $\mathtt{goto}~ \ell_{0};$\nopagebreak

\hspace{-3mm}$\ell_{h}$:~ $\mathtt{halt}$

\smallskip
\noindent
Then, this sequence of commands is translated into the following CLP facts of the form:
${\mathtt{at(}}\ell,{\textit{cmd}}{\mathtt{)}}$ meaning that at label $\ell$ there is the 
command $\mathit{cmd}$:

\smallskip

\noindent
\makebox[6mm][r]{1.~}~$\mathtt{at(0,ite(less(int(i),mult(int(2),int(n))),1,h))}$.

\noindent
\makebox[6mm][r]{2.~}~$\mathtt{at(1,ite(less(int(i),int(n)),2,4))}$.

\noindent
\makebox[6mm][r]{3.~}~$\mathtt{at(2,asgn(int(i),plus(int(i),int(1))))}$.

\noindent
\makebox[6mm][r]{4.~}~$\mathtt{at(3,goto(5))}$.

\noindent
\makebox[6mm][r]{5.~}~$\mathtt{at(4,asgn(int(j),plus(int(i),int(1))))}$.

\noindent
\makebox[6mm][r]{6.~}~$\mathtt{at(5,asgn(int(i),plus(int(i),int(1))))}$.

\noindent
\makebox[6mm][r]{7.~}~$\mathtt{at(6,goto(0))}$.

\noindent
\makebox[6mm][r]{8.~}~$\mathtt{at(h,halt)}$.

\smallskip

\noindent
We also have the following clauses that specify that an error configuration
can be reached from an initial configuration, in which case the atom {\tt incorrect}
holds:

\smallskip
\noindent
\makebox[6mm][r]{9.~}~$\mathtt{incorrect}$ {\tt{:-}} $\mathtt{initConf(X) ,\   reach(X).}$

\noindent
\makebox[6mm][r]{10.~}~$\mathtt{reach(X)}$ {\tt{:-}} $\mathtt{tr(X,X1) ,\  reach(X1).}$

\noindent
\makebox[6mm][r]{11.~}~$\mathtt{reach(X)}$ {\tt{:-}} $\mathtt{errorConf(X).}$

\smallskip
\noindent
In our case the predicates $\mathtt{initConf}$  and $\mathtt{errorConf}$
specifying the initial and the error configurations, respectively, are defined by the 
following clauses:

\smallskip
\noindent
\makebox[6mm][r]{12.~}~$\mathtt{initConf(cf(cmd(0,ite(less(int(i),mult(int(2),int(n))),1,h)),}$

\hspace{12mm}$\mathtt{[[int(i),I],[int(j),J],[int(n),N]]))}$ {\tt{:-}} $\mathtt{phiInit(I,J)}$

\noindent
\makebox[6mm][r]{13.~}~$\mathtt{errorConf(cf(cmd(h,halt),}$

\hspace{12mm}$\mathtt{[[int(i),I],[int(j),J],[int(n),N]]))}$ {\tt{:-}}
$\mathtt{phiError(I,J)}$

\noindent
\makebox[6mm][r]{14.~}~$\mathtt{phiInit(I,J)}$ {\tt{:-}} $\mathtt{I\!=\!0},\ \mathtt{J\!=\!0}.$

\noindent
\makebox[6mm][r]{15.~}~$\mathtt{phiError(I,J)}$ {\tt{:-}} $\mathtt{I\!<\!J}.$

\smallskip
\noindent
In the initial configuration (see clause~12) the command 
(labeled by~{\tt 0}) is:

\smallskip
$\mathtt{cmd(0,ite(less(int(i),mult(int(2),int(n))),1,h))}$ 

\smallskip
\noindent 
In clauses~12 and~13 the environment $\delta$ has been encoded by the list: 

\smallskip
$\mathtt{[
[int(i),I],[int(j),J],[int(n),N]]}$

\smallskip
\noindent 
which provides the bindings for the integer variables~$i,j$, and $n$, respectively.
The initial environment is any environment which binds $i$ to 0 and $j$ to 0.

The CLP program consisting of clauses 1--15 above, together with
the clauses that define the predicate {\tt tr} (see clauses {1--5} of Section~\ref{sec:syntax_semantics}), is called the {\em CLP encoding} of the given incorrectness
triple \mbox{$\bbL\varphi_{\mathit{init}}(i,j)\bbR$ {\textit{prog}} $ 
\bbL\varphi_{\mathit{error}}(i,j)\bbR$.}

\vspace*{-1mm}

\begin{theorem}{\em (Correctness of CLP Encoding)}
\label{thm:encoding}
Let $\mathit{T}$ be the CLP encoding
of the incorrectness
triple $\bbL\varphi_{\mathit{init}}\bbR $ {\textit{prog}} $
\bbL\varphi_{\mathit{error}}\bbR$.
The program
{\textit{prog}}~is correct with respect to $\varphi_{\mathit{init}}$ 
and $\varphi_{\mathit{error}}$
 iff $\,{\tt incorrect}\notin\!M(\textit{T\/})$.
\end{theorem}

\section{The Verification Method}
\label{sec:verification_method}
In this section we present a method for verifying whether or not 
a program \textit{prog} is correct with respect to any given pair of
$\varphi_{\mathit{init}}$ and $\varphi_{\mathit{error}}$ formulas.
By Theorem~\ref{thm:encoding}, this verification task
is equivalent to checking whether or not $\,{\tt incorrect}\notin\!M(\textit{T\/})$,
where $\mathit{T}$ is the CLP encoding
of the incorrectness
triple \mbox{$\bbL\varphi_{\mathit{init}}\bbR $ {\textit{prog}} $
\bbL\varphi_{\mathit{error}}\bbR$.}

Our verification method is based on transformations of CLP
programs that preserve the least model semantics~\cite{EtG96,Fi&04a}.
It makes use of the following {\em transformation rules},
collectively called {\em unfold/fold rules}: {\em unfolding}, {\em goal replacement}, 
{\em clause removal}, {\em definition}, 
and {\em folding}.
The verification method is an extension of the method
for proving safety of imperative programs by specialization of CLP
programs presented in~\cite{De&13a}.
Actually, the transformations presented in this paper are much more powerful
than program specialization and, as we will show in the next sections, they enable the
verification of correctness properties which are more complex than those 
considered in~\cite{De&13a}.

\smallskip
Our verification method consists of the following two steps. 

\noindent
{\it Step}~(A): {\em Removal of the Interpreter.}
The CLP program $T$ which encodes the incorrectness triple
\mbox{$\bbL\varphi_{\mathit{init}}\bbR $ {\textit{prog}} $
\bbL\varphi_{\mathit{error}}\bbR$} is specialized 
 with respect to the clauses encoding
 $\varphi_{\mathit{init}}$, \textit{prog}, and $\varphi_{\mathit{error}}$.
The result of this first transformation step is a new CLP program~$T_A$ 
such that ${\tt incorrect} \in M(T)$ iff ${\tt incorrect} \in M(T_A)$. 
Program~$T_A$ incorporates the operational semantics
of the imperative program~${\mathit{prog}}$, but
the clauses defining the predicate {\tt tr}, that is, the interpreter of
the imperative language we use, do not occur in $T_A$, hence the name of this transformation step.
Step~(A) is common to other verification techniques 
which are based on program specialization~\cite{De&13b,Pe&98}.

\smallskip
\noindent
{\it Step}~(B):~{\em Propagation of the initial and error properties.}
By applying the unfold/fold transformation rules, program~$
T_A$ is transformed into a new CLP program $T_B$
such that ${\tt incorrect} \in M(T_A)$ iff ${\tt incorrect} \in M(T_B)$.
This transformation exploits the interactions among the predicates encoding
the initial property  $\varphi_{\mathit{init}}$,
the operational semantics of \textit{prog},
and the error property $\varphi_{\mathit{error}}$, with the objective of deriving a program $T_B$ 
where the predicate {\tt incorrect} is defined either by (i) the fact
`{\tt incorrect.}', or by (ii) the empty set of clauses 
(that is, no clauses
for {\tt incorrect} occur in~$T_B$).
In Case (i) the imperative program \textit{prog} is incorrect
with respect to the given $\varphi_{\mathit{init}}$ and 
$\varphi_{\mathit{error}}$ properties,
while in Case (ii)  \textit{prog} is correct with respect to these properties.
There is a third case where neither (i) nor (ii) holds, that is,  
in program $T_B$ the predicate {\tt incorrect} is defined by a non-empty set of clauses
not containing the fact `{\tt incorrect.}'. In this third case we cannot conclude anything
about the correctness of~\textit{prog} by a simple inspection of~$T_B$ and, 
similarly to what 
has been proposed in~\cite{De&13a}, we iterate Step~(B)
by propagating at each iteration the initial and
error properties (that, in general, could have been modified during
the previous iterations), in the hope of
deriving a program where either Case~(i) or Case~(ii) holds. 
Obviously, due to undecidability limitations, 
we may never get to one of these two cases.

Note that either Case~(i) or Case~(ii) may hold for the program~$T_A$
that we have derived at the end of Step~(A) and, if this happens, we need not perform 
Step~(B) at all.

\smallskip

Step~(A) and Step~(B) are both instances of the {\it Transform} strategy presented
in Figure~\ref{fig:transf_proc}. These instances are obtained by suitable choices of the
{\em unfolding}, {\em generalization}, and {\em goal replacement}
auxiliary strategies.
Step~(A) has been fully automated using the MAP system~\cite{MAP} and always returns a program
with no residual call to the predicate {\tt tr} (this is a consequence of the 
fact that {\tt tr} has no recursive calls, and hence all calls to {\tt tr} can be fully unfolded).
A detailed description of Step~(A) for a simple \mbox{C-like} imperative language without
arrays can be found in~\cite{De&13a}.
As discussed below, we can also design the unfolding, generalization,
and goal replacement auxiliary strategies
in such a way that Step~(B) always terminates (see, in particular,
Theorem~\ref{thm:term_corr_Transform}). However, the design of 
auxiliary strategies that are effective in practice
is a very hard task.  Some of those strategies have been automated and, at the moment, 
 we are performing experiments for their evaluation.
 \medskip

\begin{figure}[!ht]
\noindent\hrulefill

\noindent \emph{Input\/}: Program~$P$.\\ 
\noindent \emph{Output\/}: Program $\textit{TransfP}$ such that 
$\texttt{incorrect}\!\in\! M(P)$ iff $\texttt{incorrect}\!\in\! M(\textit{TransfP})$.

\vspace*{-2mm}
\rule{30mm}{0.1mm}

\noindent \textsc{Initialization}:

$\textit{TransfP}:=\emptyset$;
~ ~~$\textit{InDefs}:=\{\texttt{incorrect:-\,c\!,\,G}\}$;
~ ~~$\textit{Defs}:= \textit{InDefs}$;

\medskip

\noindent \textit{while}~in \textit{InDefs} there is a clause~$C$
~\textit{do}

\smallskip

\hspace*{3mm}\begin{minipage}{154mm} 

\noindent \textsc{Unfolding}:


\noindent $\textit{TransfC} :=\textit{Unf\/}(C,A)$,~
where $A$ is the leftmost  atom with high predicate in the body of~$C$;

\smallskip

\noindent
\makebox[9mm][l]{\textit{while}}in $\textit{TransfC}$ there is a clause $D$
whose body contains an occurrence of an unfoldable atom $B$
~\textit{do} ~~$\textit{TransfC} := (\textit{TransfC}- \{ D\}) \cup \textit{Unf\/}(D,B)$
\hangindent=8mm 

\noindent \textit{end-while};

\smallskip

\noindent \textsc{Goal Replacement}:

\noindent
select a subset $R $  of $ \textit{TransfC\/}$; 

\noindent {\it{for every}}  $D\! \in\! R$ {\it do}

\hspace*{8mm}\makebox[4mm][l]{\it if}
\makebox[5mm][r]{(i)~} the constrained goal $\mathtt{c_1, G_1}$ occurs in
the body of $D$, 

\hspace*{13mm}\makebox[5mm][r]{(ii)~} all predicates in $\mathtt{G_1}$ are low,
~and~

\hspace*{13mm}\makebox[5mm][r]{(iii)~} $M(P)\models \forall\, ({\mathtt{c_1, G_{1}}}\!\leftrightarrow\!{\mathtt{c_2, G_{2}}})$, 

\hspace*{8mm}{\it then} ~replace  
$\mathtt{c_1, G_1}$ by $\mathtt{c_2, G_2}$ in the body of $D$;

\smallskip

\noindent \textsc{Clause Removal}:

\noindent \makebox[8mm][l]{\textit{while}}~in $\textit{TransfC}$ there is a clause
 $F$ whose body contains an unsatisfiable constraint
~\textit{do}\\
$\textit{TransfC} := \textit{TransfC}- \{ F \}$
\hangindent=8mm

\noindent \textit{end-while};

\smallskip

\noindent {\textsc{Definition} \& \textsc{Folding}:}

\noindent
\makebox[8mm][l]{\textit{while}}~in $\textit{TransfC}$ there is a clause
$E$  of the form: 
~$\texttt{H\,:-\,e,\,L,\,Q,\,R}$~ such that at least one high 
predicate occurs in $\mathtt{Q}$~ \textit{do}\hangindent=0mm

\smallskip

\hspace*{4mm}
\begin{minipage}{150mm} 

\noindent
\hangindent=9mm\hspace*{4mm}\makebox[4mm][l]{\textit{if}}~in $\textit{Defs}$ there is a clause $D$ of the form:
$\texttt{Newd\,:-\,d,\,D}$,
where:\\{\hspace*{1mm}}(i) for some substitution $\vartheta$,
$\mathtt{Q=D\,\vartheta}$, ~and~ \\
(ii) $\texttt{e}\sqsubseteq\,\texttt{d}\,\vartheta$

\noindent
\hspace*{4mm}\textit{then}~~$\textit{TransfC} :=(\textit{TransfC} -\{E \})\cup \{\texttt{H\,:-\,e,\,L,\,Newd} \,\vartheta\texttt{,\,R}\}$;

\noindent
\hspace*{4mm}\makebox[7mm][l]{\textit{else}}~~let 
$\textit{Gen}(E,\textit{Defs})$ be
~$\texttt{Newg\,:-\,g,\,G}$, ~~where:

\hspace*{13mm}\makebox[7mm][r]{(i)~}~\texttt{Newg} is an atom with 
high predicate symbol not occurring in\,$P\cup \textit{Defs}$, 

\hspace*{13mm}\makebox[7mm][r]{(ii)~}~for some substitution $\vartheta$, $\mathtt{Q} = \mathtt{G}\,\vartheta$, ~and~

\hspace*{13mm}\makebox[7mm][r]{(iii)~}~${\tt e} \sqsubseteq{\tt g}\,\vartheta$; 

\makebox[13mm]{~}$\textit{Defs}:=\textit{Defs}\cup \{\textit{Gen}(E,\textit{Defs})\}$;
\makebox[5mm]{~}$\textit{InDefs}:=\textit{InDefs} \cup \{\textit{Gen}(E,\textit{Defs})\}$;
\\
\makebox[13mm]{~}$\textit{TransfC} :=(\textit{TransfC} -\{E \})\cup 
\{\texttt{H\,:-\,e,\,L,\,Newg}\,\vartheta\texttt{,\,R}\}$
\end{minipage} 

\smallskip
\noindent
\textit{end-while};

\smallskip

\noindent $\textit{InDefs}:=\textit{InDefs}-\{C\}$;
~~~ $\textit{TransfP}:=\textit{TransfP}\cup \textit{TransfC}$;

\end{minipage} 

\smallskip

\noindent \textit{end-while};

\smallskip

\noindent {\textsc{Removal of Useless Clauses}:}

Remove from $\textit{TransfP}$ all clauses whose head predicate is useless.

\vspace*{-1mm}
\noindent\hrulefill
\vspace*{-1mm}
\caption {The \textit{Transform} strategy.}
\label{fig:transf_proc}
\vspace*{-4mm}
\end{figure}
Let us briefly illustrate  the \textit{Transform} strategy.
We assume that the CLP program $P$ taken as input by the strategy
contains a single clause defining the predicate \texttt{incorrect} of the form:
$\texttt{incorrect\,:-\,c\!,\,G}$,
where 
$\texttt{c}$ is a   
constraint (possibly $\texttt{true}$) and $\texttt{G}$ is a non-empty
conjunction of atoms. 
(The strategy can easily be extended to the case 
where  \texttt{incorrect} is defined by more than one clause.)
In particular, we will consider the program~$P$ 
made out of: (i)~the clauses defining the predicates \texttt{incorrect}
and \texttt{reach} (see {clauses~9--11} of
Section~\ref{sec:translating-correctness-into-CLP}), 
(ii)~the clauses defining the predicate~{\tt tr} (see {clauses~1--5} of
Section~\ref{sec:syntax_semantics}) which is the interpreter 
of our imperative language,  (iii)~the clauses for the predicates
 \texttt{initConf} and \texttt{errorConf}, and (iv) the clauses
defining the predicates on which~{\tt tr} depends
(such as {\tt at} and {\tt eval}). In~$P$  the predicate
\texttt{incorrect} is defined by the single clause
$\texttt{incorrect\,:-\,\texttt{initConf(X),\,reach(X)}}$.

The set of predicate symbols is partitioned 
into two subsets, called the {\em high} and {\em low} predicates,
respectively, such that no low predicate depends on a high predicate.
Recall that a predicate~{\tt p} {\em immediately depends on} a 
predicate~{\tt q} if in the program there is a clause
of the form $\mathtt{p(\ldots)\,\texttt {:-}\,\ldots,q(\ldots),\ldots}$ The {\em depends on}
relation is the transitive closure of the {\em immediately depends on} relation.

The predicates \texttt{incorrect}, \texttt{initConf}, \texttt{reach},
and \texttt{errorConf} are high predicates
and, in general, the body of the clause $\texttt{incorrect\,:-\,c\!,\,G}$ has
at least one occurrence of a high predicate.
Moreover, every new predicate introduced during the 
{\textsc{Def\-inition}} \& \textsc{Folding} phase of the
\textit{Transform} strategy is a high predicate.
This partition is needed for
guaranteeing the correctness of the {\it{Transform}}  strategy
(see Theorem~\ref{thm:term_corr_Transform}  below).

The  \textit{Transform} strategy makes use of two functions, 
\textit{Unf} and \textit{Gen}, which are used for 
performing unfolding and generalization steps, respectively.

Let us first consider the function~\textit{Unf}.\nopagebreak

Given a clause $C$ of the form $\tt H\,\texttt{:-\,c,L,A,R}$, where
$\texttt{H}$ and $\texttt{A}$ are atoms, 
$\texttt{c}$ is a constraint,
and $\texttt{L}$ and~$\texttt{R}$ are (possibly empty)
conjunctions of atoms, let
$\{\texttt{K}_i\, \texttt{:-}\, \texttt{c}_i\texttt{,B}_i \mid  i=1, \ldots, m\}$ 
be the set of the (renamed apart) clauses of program~$P$ 
such that, for $i\!=\!1,\ldots ,m,$ 
$\texttt{A}$ is unifiable with $\texttt{K}_i$ via the most general 
unifier~$\vartheta_i$ and $(\texttt{c,c}_i)\, \vartheta_i$ is satisfiable 
(thus, the unfolding function performs also some constraint solving operations). 
We define the following function:\nopagebreak

\smallskip
$\textit{Unf\/}(C,\texttt{A})  = \{(\texttt{H}\, \texttt{:-}\, \texttt{c,c}_i\texttt{,L,B}_i\texttt{,R})\,\vartheta_i \mid i=1,\ldots, m\}$

\smallskip\noindent
Each clause in $\textit{Unf\/}(C,\texttt{A})$ is said to be derived by {\em unfolding $C$ 
w.r.t.}~\texttt{A}.
In order to perform unfolding steps during the execution of the {\em Transform} strategy,
we assume that it is given a {\em selection function}
that determines which atoms occurring in the body of clause~$C$
are {\em unfoldable}.
This selection function, which depends on the sequence of applications of $\textit{Unf\/}$ through which we have derived~$C$, will
ensure that any sequence of clauses constructed by unfolding
w.r.t.~unfoldable atoms is finite. A survey
of techniques which ensure the finiteness of unfolding can be
found in~\cite{LeB02}.

The function {\it{Gen}}, 
called the {\em generalization operator}, is used for introducing 
new predicate definitions.
Indeed,  given a clause $E$ and the set
\textit{Defs} of clauses defining the predicates
introduced so far,  
$\textit{Gen}(E,\textit{Defs})$ returns a new predicate definition~$G$
such that $E$ can be {\em folded} by using clause~$G$.
The generalization operator guarantees
that during the execution of the \textit{Transform} strategy, a
{\em finite} number of
new predicates is introduced (otherwise, the strategy does not terminate).
One can define several generalization operators based on the notions of 
{\em widening}, {\em convex hull},  {\em most specific generalization},
and {\em well-quasi orderings}
which have been introduced for analyzing and transforming constraint logic 
programs (see, for instance,~\cite{CoC77,CoH78,De&99,Fi&13a,PeG02}). 
For lack of space we do
not present here the definitions of those generalization operators, 
and we refer the reader to the relevant literature.

The equivalences which are considered when performing the goal replacements
are called {\em laws} and their validity in the least model~$M(P)$ 
can be proved once and for all
before applying the \textit{Transform} strategy. 
During the application of the \textit{Transform} strategy we also need
an auxiliary strategy for performing the goal replacement steps
(this auxiliary strategy has to select the clauses where the replacements 
should take place
and the law to be used).
In Section~\ref{sec:array_properties} we will consider programs acting on arrays 
and we will see an application of the goal replacement rule
which makes use of a law that holds for arrays.
However,
we leave it for future work the study of general strategies for 
performing goal replacement.

In the \textit{Transform} strategy we also use the following notions.
We say that a constraint \texttt{c} {\em entails} a constraint \texttt{d},
denoted $c \sqsubseteq d$, if $\forall\, (\mathtt{c}\! \rightarrow\! \mathtt{d})$ holds, where, 
as usual, by $\forall(\varphi)$ 
we denote the universal closure of a formula~$\varphi$.
The set of \emph{useless predicates} in a program \( P \) is the
maximal set \( U \) of predicates  occurring in \( P \) such that 
\texttt{p} is in \( U \) iff every clause with
head predicated \texttt{p} is of the form \(\mathtt{ p(\ldots)\ \texttt{:-}\ c, 
G_{1}, q(\ldots ), G_{2} }\), for some \( \texttt q \) in $U$.
A clause in a program~$P$ is {\em useless} if the predicate of its
head is useless in $P$.

We have the following result.

\vspace{-2mm}
\begin{theorem}{\em (Termination and Correctness of the \textit{Transform} strategy)}
\label{thm:term_corr_Transform} 
{\em (i)}~The Transform strategy terminates.
{\em (ii)}~Let program $\textit{TransfP}$ be the output of the Transform strategy
applied on the input program $P$.
Then, 
$\mathtt{incorrect}\! \in \!M(\textit{P\/})$ \mbox{iff}
$\mathtt{incorrect}\! \in\! M(\textit{TransfP})$.
\end{theorem}

\vspace{-1mm}
\noindent

The termination of the  \textit{Transform} strategy is guaranteed
by the assumption that one can make only a finite number of 
unfolding and generalization steps.
The correctness of the strategy
with respect to the least model semantics 
directly follows from the correctness of the transformation rules~\cite{EtG96,TaS86}.
Indeed, the conditions on high and low predicates ensure that the \textit{Transform}
strategy complies with the conditions on predicate {\it levels} 
given in~\cite{TaS86}.

\medskip

Let us briefly explain how the  \textit{Transform} strategy  may achieve the
objective of deriving a program $\textit{TransfP}$ with either the fact 
`$\texttt{incorrect.}$' (hence proving that \textit{prog} is incorrect) or the empty set of clauses for the predicate  $\texttt{incorrect}$  (hence proving that \textit{prog} is correct).

If \textit{prog} is incorrect, the fact `$\texttt{incorrect.}$'  can, in principle, be derived
by unfolding  the clause
$\texttt{incorrect\,:-\,c\!,\,G}$.
Indeed, observe that if $\texttt{incorrect}\in M(P)$ then,
by the completeness of the {\em top-down} evaluation strategy of CLP
programs~\cite{JaM94},
there exists a sequence of unfolding steps that leads
to the fact  `$\texttt{incorrect.}$'.
In practice, however, the ability to derive such a fact depends on the 
specific strategy used for selecting the atoms for performing unfolding steps
during the \textsc{Unfolding}
phase.

A program $\textit{TransfP}$  where the predicate 
$\texttt{incorrect}$ is defined by the empty set of clauses
can be obtained by first 
deriving a program where the set of clauses defining
\texttt{incorrect} and the high predicates upon
which  \texttt{incorrect} depends, contains no constrained facts.
Indeed, in this case the predicate \texttt{incorrect} 
is useless and all the clauses of its definition are removed by the last step
of the \textit{Transform} strategy.
A set of clauses without constrained facts may be derived as follows.
We perform the \textsc{Unfolding}, \textsc{Goal Replacement},
and \textsc{Clause Removal} phases as indicated in the
\textit{Transform} strategy.
If we get a set $\textit{TransfC}$ of clauses with constrained facts,
then the strategy will necessarily derive a final program $\textit{TransfP}$
with constrained facts. 
Otherwise, all clauses in $\textit{TransfC}$ are folded by (possibly) introducing new 
predicate definitions and the strategy continues by processing
the newly introduced predicates (if any).
If the strategy exits the while-loop without producing any
constrained fact, we derive a program  $\textit{TransfP}$ 
defining a set of mutually recursive predicates without any constrained fact.

\medskip

Let us  consider the incorrectness triple 
$\bbL \varphi_{\mathit{init}}(i,j) \bbR$ {\it{increase}}
$\bbL \varphi_{\mathit{error}}(i,j)\bbR$ of Section~\ref{sec:translating-correctness-into-CLP}. 
In order to show that
the program \textit{increase} is correct with respect to 
$\varphi_{\mathit{init}}=_{\mathit{def}}i\!=\!0 \andw j\!=\!0$ 
and $\varphi_{\mathit{error}}=_{\mathit{def}}i\!<\!j$,
we start off from clauses~1--15 
associated
with the program \textit{increase} (see Section~\ref{sec:translating-correctness-into-CLP}), 
together with
the clauses for the predicate~$\mathtt{tr}$ ({clauses~1--5} of  
Section~\ref{sec:syntax_semantics} and the clauses on 
which~$\mathtt{tr}$ depends)
which, as already mentioned, define the interpreter of our imperative language.

We perform Step~(A) of the verification method by applying the 
{\it{Transform}} strategy.

\smallskip

\noindent
\textsc{Unfolding.}
First we unfold
clause~9 with respect to the leftmost atom with high predicate, that is, 
$\mathtt{initConf(X)}$, and 
we get:

\smallskip

\noindent
16.~$\mathtt{incorrect}$ {\tt{:-}} $\mathtt{I\!=\!0},\  \mathtt{J\!=\!0},\ \mathtt{reach(cf(cmd(0,ite(less(int(i),mult(int(2),int(n))),1,h)),}$\nopagebreak

\hspace{60mm}$\mathtt{[[int(i),I],[int(j),J],[int(n),N]])).}$

\smallskip

\noindent
\textsc{Definition \& Folding.} We introduce the new predicate definition:

\smallskip

\noindent
17.~$\mathtt{new1(I,J,N)}$ {\tt{:-}} $\mathtt{reach(cf(cmd(0,ite(less(int(i),mult(int(2),int(n))),1,h)),}$\nopagebreak

\hspace{60mm}$\mathtt{[[int(i),I],[int(j),J],[int(n),N]])).}$

\smallskip

\noindent We fold clause 16 using clause 17 and we get:

\smallskip

\noindent
18.~$\mathtt{incorrect}$ {\tt{:-}} $\mathtt{I\!=\!0,\  J\!=\!0,\  new1(I,J,N).}$

\smallskip

\noindent Then, we continue the execution of the \textit{Transform} strategy,
starting from the last definition we have introduced, that is, clause~17
(indeed, we have ${\mathit{InDefs}}\!=\!\{\mbox{clause 17}\}$).
Eventually we get the following program~$T_A$:

\smallskip
\noindent
18.~$\mathtt{incorrect}$ {\tt{:-}} $\mathtt{I\!=\!0,~~ J\!=\!0,~~ new1(I,J,N).}$
   
\noindent
19.~$\mathtt{new1(I,J,N)}$ {\tt{:-}} $\mathtt{I\!<\!2\!*\!N,~~I\!<\!N,~~ I1\!=\!I\!+\!2,~~ \mathtt{new1(I1,J,N)}.}$

\noindent
20.~$\mathtt{new1(I,J,N)}$ {\tt{:-}} $\mathtt{I\!<\!2\!*\!N,~~I\!\geq\!N,~~ I1\!=\!I\!+\!1,~~ 
J1\!=\!I\!+\!1,~~  \mathtt{new1(I1,J1,N)}.}$

\noindent
21.~$\mathtt{new1(I,J,N)}$ {\tt{:-}} $\mathtt{I\!\geq\!2\!*\!N,~~ I\!<\!J.}$

\smallskip
\noindent
Now we have completed Step~(A). 

Unfortunately, it is not possible to prove by direct evaluation
that \texttt{incorrect} is not a consequence of the above CLP clauses.
Indeed, the  evaluation of the query \texttt{incorrect} using the standard
top-down strategy enters into an infinite loop.
Tabled evaluation~\cite{CuW00} does not terminate either, as infinitely many tabled atoms are generated. Analogously, bottom-up evaluation is 
unable to return an answer.
Indeed, the presence of a constrained fact for $\mathtt{new1}$ in  program 
$T_A$ (see clause~21) generates in the least model $M(T_A)$, by repeatedly using the recursive clauses~19 and 20, infinitely 
many new atoms with predicate ${\mathtt{new1}}$, and thus
we cannot show that ${\mathtt{new1}}$ does not hold for 
$\mathtt{I\!=\!0 \andw J\!=\!0}$. 

Hence, in this way we cannot show that
$\mathtt{incorrect}$ does not hold in $M(T_A)$ and we cannot conclude
that the program {\it{increase}} is correct. 

Our verification method, instead of directly evaluating the query \texttt{incorrect} in $T_A$,
proceeds to Step~(B).
In order to perform this transformation step
we may choose to propagate either the initial property or the error property.
Let us opt for the second choice. (However, verification would succeed
also by propagating the initial property.)
In order to do so,  we have first to `reverse' 
program~$T_A$, as indicated in~\cite{De&13a}.
We proceed as follows.
First, program~$T_A$ is transformed into a program of the form:

\smallskip
\noindent
s1.~\makebox[18mm][l]{$\mathtt{incorrect}$} {\tt{:-}} 
$\mathtt{a(U), ~r1(U).}$\nopagebreak

\noindent
s2.~\makebox[9mm][l]{$\mathtt{r1(U)}$} {\tt{:-}} 
$\mathtt{trans(U,V), ~r1(V).}$\nopagebreak

\noindent
s3.~\makebox[9mm][l]{$\mathtt{r1(U)}$} {\tt{:-}} 
$\mathtt{b(U).}$

\smallskip
\noindent 
where the predicates $\mathtt{a(U)}$, 
$\mathtt{trans(U,V)}$, and $\mathtt{b(U)}$ are defined by the 
following clauses:

\smallskip
\noindent
s4.~$\mathtt{a([new1,I,J,N])}$ {\tt{:-}} $\mathtt{I\!=\!0,~~ J\!=\!0.}$\nopagebreak 

\noindent
s5.~$\mathtt{trans([new1,I,J,N],[new1,I1,J,N])}${\tt{~:-}} 
$\mathtt{I\!<\!2\!*\!N,~~I\!<\!N,~~I1\!=\!I\!+\!2.}$ 

\noindent
s6.~$\mathtt{trans([new1,I,J,N],[new1,I1,J1,N])}${\tt{:-}} 
$\mathtt{I\!<\!2\!*\!N,~~I\!\geq\!N,~~I1\!=\!I\!+\!1,~~J1\!=\!I\!+\!1.}$

\noindent
s7.~$\mathtt{b([new1,I,J,N])}$ {\tt{:-}} $\mathtt{I\!\geq\!2\!*\!N,~~ I\!<\!J.}$

\smallskip
\noindent
This transformation is correct because program~$T_A$ can be obtained from clauses~s1--\,s7
by: (i)~unfolding  clauses~s1--\,s3 with respect to~$\tt a(U)$, $\tt trans(U,V)$, and
$\tt b(U)$, and then (ii)~rewriting the atoms
of the form $\mathtt{r1([new1,X,Y,N])}$ 
as $\mathtt{new1(X,Y,N)}$.
(The occurrences of the predicate symbol
$\mathtt{new1}$ in the arguments
of $\mathtt{r1}$ should be considered as an individual constant.)
Then, the {\rm{reversed program}}~$T_A^{\textit{rev}}$
is given by the following clauses (with the same definitions of $\mathtt{a(U)}$, 
$\mathtt{trans(U,V)}$, and $\mathtt{b(U)}$):

\smallskip

\noindent
rev1.~~\makebox[17mm][l]{$\mathtt{incorrect}$} {\tt{:-}} $\mathtt{b(U), 
~r2(U).}$\nopagebreak

\noindent 
rev2.~~\makebox[9mm][l]{$\mathtt{r2(V)}$} {\tt{:-}} 
$\mathtt{trans(U,V), ~r2(U).}$\nopagebreak

\noindent
rev3.~~\makebox[9mm][l]{$\mathtt{r2(U)}$} {\tt{:-}} 
$\mathtt{a(U).}$

\smallskip
\noindent
One can show that program reversal is correct in the sense that
$\mathtt{incorrect}\in M(T_A)$ iff $\mathtt{incorrect} \in 
M(T_A^{\textit{rev}})$~\cite{De&13a}. 

Now, we perform Step~(B) of the verification method.
Let us apply the \textit{Transform} strategy taking as input 
program $T_A^{\textit{rev}}$ (clauses rev1--rev3) and clauses s4--\,s7. 
We assume that the high predicates are: {\tt{incorrect}}, {\tt{b}}, {\tt{r2}}, {\tt{trans}},
and {\tt{a}}.

\smallskip

\noindent
\textsc{Unfolding.}
First we unfold clause rev1 w.r.t.~the leftmost atom with high predicate, that is, {\tt b(U)},
 and we get:

\smallskip
\noindent
22.~$\mathtt{incorrect}$ {\tt{:-}} 
$\mathtt{I\!\geq\!2\!*\!N,~~ I\!<\!J,~~ r2([new1,I,J,N]).}$

\smallskip

\noindent  The {\sc goal replacement} and 
the {\sc clause removal} phases leave unchanged the set of clauses 
we have derived so far.

\smallskip

\noindent  \textsc{Definition \& Folding.} In order to fold clause 22 we introduce the clause:

\smallskip
\noindent
23.~$\mathtt{new2(I,J,N)}$ {\tt{:-}} 
$\mathtt{I\!\geq\!2\!*\!N,~~ I\!<\!J,~~ r2([new1,I,J,N]).}$

\smallskip
\noindent
and we fold clause 22 using clause 23. Thus, we get:

\smallskip
\noindent
24.~$\mathtt{incorrect}$ {\tt{:-}} 
$\mathtt{I\!\geq\!2\!*\!N,~~  I\!<\!J,~~ new2(I,J,N).}$

\smallskip
\noindent
At this point we execute again the outermost body of the while-loop of the \textit{Transform}
strategy because {\it{InDefs}} contains clause~23, which is 
not a constrained fact. 

\smallskip

\noindent \textsc{Unfolding.} By unfolding clause~23 w.r.t.~the 
atom $\mathtt{r2([new1,I,J,N])}$, we get the following two clauses:

\smallskip
\noindent
25.~$\mathtt{new2(I,J,N)}$ {\tt{:-}} $\mathtt{I\!\geq\!2\!*\!N,~~  
I\!<\!J,~~ trans(U,[new1,I,J,N]),~~ r2(U).}$ 

\noindent
26.~$\mathtt{new2(I,J,N)}$ {\tt{:-}} $\mathtt{I\!\geq\!2\!*\!N,~~  
I\!<\!J,~~ a([new1,I,J,N]).}$

\smallskip
\noindent
Then, by unfolding clause~25 w.r.t.~the atom~$\mathtt{trans(U,[new1,I,J,N])}$, 
we get:

\smallskip
\noindent
27.~$\mathtt{new2(I1,J,N)}$ {\tt{:-}} $\mathtt{I1\!=\!I\!+\!2,~~ 
I\!<\!2\!*\!N,~~ I\!<\!N,~~ I\!+\!2\!\geq\!2\!*\!N,~~ I\!+\!2\!<\!J,~~ r2([new1,I,J,N]).}$ 

\smallskip
\noindent
By unfolding clause~26 w.r.t.~$\mathtt{a([new1,I,J,N])}$,
we get an empty set of clauses  (indeed, the constraint 
`$\mathtt{I\!<\!J, I\!=\!0, J\!=\!0}$' is unsatisfiable). 

\smallskip

\noindent
\textsc{Definition \& Folding.} 
In order to fold clause~27 we perform a generalization operation as follows.
Clause~27 can be folded introducing the clause:

\smallskip\noindent
28.~$\mathtt{new3(I,J,N)}$ {\tt{:-}} $\mathtt{I\!<\!2\!*\!N,~~ 
I\!<\!N,~~ I\!+\!2\!\geq\!2\!*\!N,~~ I\!+\!2\!<\!J,~~ r2([new1,I,J,N]).}$ 

\smallskip\noindent
However, the comparison between 
clause~23 introduced in a previous step and clause~28 
shows the risk of introducing 
an unlimited number of definitions whose body contains the atom $\mathtt{r2([new1,I,J,N])}$
(see, in particular, 
the constraint `$\mathtt{I\!\geq\!2\!*\!N,\ I\!<\!J}$'
in clause~23 and the constraint 
`$\mathtt{I\!+\!2\!\geq\!2\!*\!N,\ I\!+\!2\!<\!J}$' in clause~28),
thereby making the {\it{Transform}} strategy diverge. 
To avoid this divergent behaviour, we apply {\em widening}~\cite{CoH78} to
 clauses~23 and~28, and
we introduce the following clause~29, instead of clause~28:

\smallskip
\noindent
29.~$\mathtt{new3(I,J,N)}$ {\tt{:-}} $\mathtt{I\!<\!J,~~ r2([new1,I,J,N]).}$

\smallskip
\noindent
Recall that the widening operator applied to two clauses $c1$ and $c2$ (in this order)
behaves, in general, as follows. After replacing every equality constraint
$\mathtt{A\!=\!B}$ by the equivalent conjunction `$\mathtt{A\!\leq\! B,~ A\!\geq\! B}$',
the widening operator
 returns a clause which is like~$c1$, except that the atomic constraints are
only those of~$c1$ which are implied by the constraint of~$c2$.
In our case, the widening operator drops the atomic constraint 
$\mathtt{I\!\geq\!2\!*\!N}$ and keeps $\mathtt{I\!<\!J}$ only.

\smallskip
\noindent
By folding clause 27 w.r.t.~the atom $\mathtt{r2([new1,I,J,N])}$, we get:

\smallskip
\noindent
30.~$\mathtt{new2(I1,J,N)}$ {\tt{:-}} $\mathtt{I1\!=\!I\!+\!2,~~
I\!<\!2\!*\!N,~~ I\!<\!N,~~ I\!+\!2\!\geq\!2\!*\!N,~~ I\!+\!2\!<\!J,~~ new3(I,J,N).}$

\smallskip
\noindent
Now, we process the newly introduced definition clause~29 and
we perform a new iteration of the body of the outermost while-loop of
the \textit{Transform} strategy.

\smallskip

\noindent \textsc{Unfolding.} After a few unfolding steps from clause~29, we get:

\smallskip
\noindent
31.~$\mathtt{new3(I1,J,N)}$ {\tt{:-}} 
$\mathtt{I1\!=\!I\!+\!2,~~ 
I\!<\!2\!*\!N,~~ I\!<\!N,~~ I\!+\!2\!<\!J,~~ r2([new1,I,J,N])}$.

\smallskip
\noindent
\textsc{Definition \& Folding.} 
In order to fold clause~31 we do not 
need to introduce any new definition. 
Indeed, it is possible to fold clause~31 by using clause 29
and we get:

\smallskip
\noindent
32.~$\mathtt{new3(I1,J,N)}${\tt{:-}}$\mathtt{I1\!=\!I\!+\!2,~~ 
I\!<\!2\!*\!N,~~ I\!<\!N,~~ I\!+\!2\!<\!J,~~ new3(I,J,N)}$.

\smallskip
\noindent
Now, \textit{InDefs} is empty and we exit the outermost while-loop. We get the following program~$T_B$:

\smallskip
\noindent
24.~$\mathtt{incorrect}$ {\tt{:-}} $\mathtt{I\!\geq\!2\!*\!N,~~  
I\!<\!J,~~ new2(I,J,N).}$

\noindent
30.~$\mathtt{new2(I1,J,N)}$ {\tt{:-}} $\mathtt{I1\!=\!I\!+\!2,~~ 
I\!<\!2\!*\!N,~~ I\!<\!N,~~ I\!+\!2\!\geq\!2\!*\!N,~~ I\!+\!2\!<\!J,~~ new3(I,J,N).}$ 

\noindent
32.~$\mathtt{new3(I1,J,N)}$ {\tt{:-}} 
$\mathtt{I1\!=\!I\!+\!2,~~ 
I\!<\!2\!*\!N,~~ I\!<\!N,~~ I\!+\!2\!<\!J,~~ new3(I,J,N)}$.

\smallskip
\noindent
Since program~$T_B$ contains no constrained facts,
all its clauses are useless and can be removed from~$T_B$.
Thus, at the end of Step~(B) we get the final empty program~$T_B$.
Hence, $M(T_B)$ is the empty set and
the atom $\mathtt{incorrect}$ does not belong to it.
We conclude that the given program {\it{increase}} is
correct with respect to the given properties $\varphi_{\mathit{init}}$ and 
$\varphi_{\mathit{error}}$.

The various transformation steps which lead to program $T_B$,
including the removal of the interpreter, the program reversal,
and the generalization steps performed during Step~(B),
have been automatically
performed by the MAP system~\cite{MAP} (see~\cite{De&13a}
for some experimental results).

\section{Verifying Complex Correctness Properties by Conjunctive Folding}
\label{sec:conjunctivefolding}

In this section we show how our verification method can be used 
also in the case when the error properties are specified by sets of
CLP clauses, rather than by constraints only.
In order to deal with this class of error properties
we make use of
transformation rules which are more powerful than the ones used in the
verification example of the previous section.
Indeed, during the \textsc{Definition\&Folding} phase
of the \textit{Transform} strategy, we allow ourselves to 
introduce new predicates by using definition clauses of the form:

$\texttt{Newp}\, \texttt{:-}\, \mathtt{c,\, G}$

\noindent
where \texttt{Newp} is an atom with a new predicate symbol,
\texttt{c} is a constraint, and
\texttt{G} is a {\em conjunction of one or more atoms}. Clauses of 
that form will then be used for performing folding steps.
(Note that the new predicate definitions introduced during the verification
example of the previous section 
are all of the form:~ $\texttt{Newp}\, \texttt{:-}\, \mathtt{c,\, A}$, 
where \texttt{A} is a single atom.)
The more powerful definition and folding rules,
called {\em conjunctive definition} and {\em conjunctive folding},
respectively, allow us to perform verification tasks
that cannot be handled by
the technique presented in~\cite{De&13a}, which is based on {\em atomic} definition and {\em atomic} folding.

\smallskip

Let us consider the following program \textit{gcd} 
for computing the {\em greatest common divisor}~$z$
of two positive integers $m$ and $n$:

\hspace{-4mm}$\ell_{0}$: $x=m\,;$

\hspace{-4mm}$\ell_{1}$: $y=n\,;$

\hspace{-4mm}$\ell_{2}$: $\mathtt{while}~(x\neq y)\ 
\{\,\mathtt{if}~(x>y)~x\!=\!x\!-\!y\,;~\mathtt{else}~y\!=\!y\!-\!x\,\}$\,;

\hspace{-4mm}$\ell_{3}$: $z=x\,;$

\noindent
\makebox[5mm][l]{} 
\hspace{-4mm}$\ell_{h}$: $\mathtt{halt}$

\noindent
and the incorrectness triple 
$\bbL\varphi_{\mathit{init}}(m,n)\bbR \ \textit{gcd\/}\ 
\bbL\varphi_{\mathit{error}}(m,n,z)\bbR$,
where:

\noindent
-- $\varphi_{\mathit{init}}(m,n)$ is 
$ m\!\geq\!1 \andw n\!\geq\!1$, and \nopagebreak

\noindent
--  $\varphi_{\mathit{error}}(m,n,z)$ is  $\exists\, d\,(\textit{gcd\/}(m,n,d\/) 
\andw d\!\neq\! z)$. 
The~property $\varphi_{\mathit{error}}$  uses
the ternary predicate~\textit{gcd} defined by the following CLP clauses:

\smallskip
\noindent
\makebox[5mm][l]{1.~}$\mathtt{gcd(X,Y,D)~ \texttt{:-} ~X\!>\!Y,~~ X1\!=\!X\!-\!Y,~~ gcd(X1,Y,D).}$

\noindent
\makebox[5mm][l]{2.~}$\mathtt{gcd(X,Y,D)~ \texttt{:-} ~X\!<\!Y,~~ Y1\!=\!Y\!-\!X,~~ gcd(X,Y1,D).}$

\noindent
\makebox[5mm][l]{3.~}$\mathtt{gcd(X,Y,D)~ \texttt{:-} ~X\!=\!Y,~~ Y\!=\!D.}$

\smallskip
\noindent
Thus, the incorrectness triple holds if and only if, for some positive integers
$m$  and $n$, the program \textit{gcd}
computes a value of $z$ that is different from the greatest common divisor of 
$m$ and $n$.

As indicated in Section~\ref{sec:verification_method}, the program \textit{gcd} 
can be translated into a set of CLP facts defining the predicate~\texttt{at}. We will not show them here.

The predicates $\mathtt{initConf}$  and $\mathtt{errorConf}$
specifying the initial and the error configurations, respectively, are defined by the 
following clauses:

\smallskip
\noindent
\makebox[5mm][l]{4.~}$\mathtt{initConf(cf(cmd(0,asgn(int(x),int(m))),}$

\hspace{8mm}$\mathtt{[[int(m),M],[int(n),N],[int(x),X],[int(y),Y],[int(z),Z]]))}$ {\tt{:-}}~ $\mathtt{phiInit(M,N).}$

\noindent
\makebox[5mm][l]{5.~}$\mathtt{errorConf(cf(cmd(h,halt),}$

\hspace{8mm}$\mathtt{[[int(m),M],[int(n),N],[int(x),X],[int(y),Y],[int(z),Z]]))}$ {\tt{:-}} ~$\mathtt{phiError(M,N,Z).}$

\noindent
\makebox[5mm][l]{6.~}$\mathtt{phiInit(M,N)}$ {\tt{:-}} $\mathtt{M\!\geq\!1},~~ \mathtt{N\!\geq\!1}.$

\noindent
\makebox[5mm][l]{7.~}$\mathtt{phiError(M,N,Z)}$ {\tt{:-}} $\mathtt{gcd(M,N,D),~~ D \!\neq\! Z}.$

\smallskip
\noindent
The  CLP program encoding the given incorrectness triple 
consists of clauses~1--7 above, together with the clauses defining the 
predicate \texttt{\,at\,} that encode the program~\textit{gcd}, and the clauses
that define the predicates {\tt incorrect},  {\tt reach}, and {\tt tr} (that is, clauses~9--11 of
Section~\ref{sec:translating-correctness-into-CLP} and clauses~1--5 of Section~\ref{sec:syntax_semantics}).

Now we perform Step~(A) of our verification method, which consists in
the removal of the interpreter, and we derive the following CLP program:

\smallskip

\noindent
\makebox[6mm][r]{8.\hspace{1.5mm}}$\mathtt{incorrect\ \texttt{:-}\ M\!\geq\!1,~~ N\!\geq\! 1,~~ new1(M,N,M,N,Z).}$

\noindent
\makebox[6mm][r]{9.\hspace{1.5mm}}$\mathtt{new1(M,N,X,Y,Z)\ \texttt{:-}\ X\!>\!Y,~~ X1\!=\!X\!-\!Y,~~ new1(M,N,X1,Y,Z).}$

\noindent
\makebox[6mm][l]{10.}$\mathtt{new1(M,N,X,Y,Z)\ \texttt{:-}\  X\!<\!Y,~~ Y1\!=\!Y\!-\!X,~~ new1(M,N,X,Y1,Z).}$

\noindent
\makebox[6mm][l]{11.}$\mathtt{new1(M,N,X,Y,Z)\ \texttt{:-}\  X\!=\!Y,~~ Z\!=\!X,~~ gcd(M,N,D),~~ Z\!\neq\!D.}$

\smallskip

\noindent
Clauses 8 and 11 can be rewritten, respectively, as:

\smallskip

\noindent
\makebox[7mm][r]{8r.~}$\mathtt{incorrect\ \texttt{:-}~ M\!\geq\!1,~~ N\!\geq\! 1,~~ Z\!\neq\!D,~~
new1(M,N,M,N,Z),~~ gcd(M,N,D).}$ 

\noindent
\makebox[7mm][r]{11r.~}$\mathtt{new1(M,N,X,Y,Z)\ \texttt{:-}\  X\!=\!Y,~~ Z\!=\!X.}$

\smallskip

\noindent
This rewriting is correct because {\tt{new1}} modifies the values of neither {\tt M} nor {\tt N}.

Note that we could avoid performing the above rewriting and automatically
derive a similar program 
where the constraints characterizing the initial and the error properties 
occur in the same clause,
by starting our derivation from a more general definition of the reachability relation.
However, an in-depth analysis of this variant of our verification method
is beyond the scope of this paper (see also \cite{Pe&98}
for a discussion about different styles of encoding 
the reachability relation and the semantics of imperative languages in CLP).

Now we perform Step (B) of the verification method by applying the \textit{Transform}
strategy to the program consisting of clauses $\{1,2,3,8{\rm r},9,10,11{\rm r}\}$.
We assume that the only high predicates are {\tt{incorrect}} and {\tt new1}.

\smallskip

\noindent
\textsc{Unfolding.}
We start off by unfolding clause 8r w.r.t.~the atom $\mathtt{new1(M,N,M,N,Z)}$,
and we get: 

\smallskip

\noindent
\makebox[6mm][r]{12.~}$\mathtt{incorrect \ \texttt{:-}\ M\!\geq\! 1,~~ N\!\geq\! 1,~~  M\!>\!N,~~ X1\!=\!M\!-\!N,~~ Z\!\neq\!D,~}$ 
$\mathtt{new1(M,N,X1,N,Z),~~ gcd(M,N,D).}$

\noindent
\makebox[6mm][r]{13.~}$\mathtt{incorrect\ \texttt{:-}\ M\!\geq\! 1,~~ N\!\geq\! 1,~~ M\!<\!N,~~ Y1\!=\!N\!-\!M,~~ Z\!\neq\!D,~}$
$\mathtt{new1(M,N,M,Y1,Z),~~ gcd(M,N,D).}$

\noindent
\makebox[6mm][r]{14.~}$\mathtt{incorrect \ \texttt{:-}\ M\!\geq\! 1,~~ N\!\geq\! 1,~~ 
M\!=\!N,~~ Z\!=\!M,~~ Z\!\neq\!D,~~ gcd(M,N,D).}$ 

\smallskip

\noindent
By unfolding clauses 12--14 w.r.t.~the atom $\mathtt{gcd(M,N,D)}$ we derive:

\smallskip

\noindent
\makebox[6mm][r]{15.~}$\mathtt{incorrect \ \texttt{:-}\ M\!\geq\! 1,~~ N\!\geq\! 1,~~ M\!>\!N,~~ X1\!=\!M\!-\!N,~~ Z\!\neq\!D,~~ new1(M,N,X1,N,Z),~~  gcd(X1,N,D).}$

\noindent
\makebox[6mm][r]{16.~}$\mathtt{incorrect \ \texttt{:-}\ M\!\geq\! 1,~~ N\!\geq\! 1,~~ M\!<\!N,~~ Y1\!=\!N\!-\!M,~~ Z\!\neq\!D,~~ new1(M,N,M,Y1,Z),~~ gcd(M,Y1,D).}$

\smallskip
\noindent
(Note that by unfolding clause~14 we get an empty set of clauses because 
the constraints
derived in this step are all unsatisfiable.)

\smallskip
\noindent
The \textsc{Goal Replacement} and \textsc{Clause Removal} phases 
do not modify the set of clauses derived after the \textsc{Unfolding} phase
because no laws are available for the predicate \texttt{gcd}.

\smallskip
\noindent
\textsc{Definition \& Folding.} 
In order to fold clauses 15 and 16, we perform a generalization step
and we introduce a new predicate defined by the following clause:

\smallskip

\noindent
\makebox[6mm][r]{17.~}$\mathtt{new2(M,N,X,Y,Z,D) \ \texttt{:-}\ M\!\geq\! 1,~~ N\!\geq\! 1,~~ 
 Z\!\neq\!D,~~ new1(M,N,X,Y,Z),~~ gcd(X,Y,D).}$

\smallskip\noindent
The body of this clause is the most specific generalization of the 
bodies of clauses~8r,~15 and~16. Here, we define a conjunction $G$ to be
a generalization of a conjunction $C$ if there exists a substitution $\vartheta$
such that $G\vartheta$ can be obtained by deleting some of the conjuncts of $C$.

Now, clauses 15 and 16 can be folded by using clause~17, thereby deriving:\nopagebreak

\smallskip

\noindent
\makebox[6mm][r]{18.~}$\mathtt{incorrect \ \texttt{:-}\ M\!\geq\! 1,~~ N\!\geq\! 1,~~ M\!>\!N,~~ X1\!=\!M\!-\!N,~~ Z\!\neq\!D,~~ new2(M,N,X1,N,Z,D).}$\nopagebreak


\noindent
\makebox[6mm][r]{19.~}$\mathtt{incorrect \ \texttt{:-}\ M\!\geq\! 1,~~ N\!\geq\! 1,~~ M\!<\!N,~~ Y1\!=\!N\!-\!M,~~ Z\!\neq\!D,~~ new2(M,N,M,Y1,Z,D).}$

\medskip\noindent
Clause~17 defining the new predicate \texttt{new2}, is added to \textit{Defs} and 
\textit{InDefs} and, since the latter set is not empty, 
we perform a new iteration of the while-loop body of the \textit{Transform} strategy.
 
\smallskip

\noindent\textsc{Unfolding.}
By unfolding clause 17 
w.r.t.~\texttt{new1(M,N,X,Y,Z)} and then unfolding the resulting clauses
w.r.t.~\texttt{gcd(X,Y,Z)}, we derive:

\smallskip

\noindent
\makebox[6mm][r]{20.~}$\mathtt{new2(M,N,X,Y,Z,D) \ \texttt{:-}\ M\!\geq\! 1,~~ N\!\geq\! 1,~~ X\!>\!Y,~~ X1\!=\!X\!-\!Y,~~ Z\!\neq\!D,~~ }$ $\mathtt{new1(M,N,X1,Y,Z),~~ gcd(X1,Y,D).}$

\noindent
\makebox[6mm][r]{21.~}$\mathtt{new2(M,N,X,Y,Z,D) \ \texttt{:-}\ M\!\geq\! 1,~~ N\!\geq\! 1,~~ X\!<\!Y,~~ Y1\!=\!Y\!-\!X,~~ Z\!\neq\!D,~~ }$ $\mathtt{new1(M,N,X,Y1,Z),~~ gcd(X,Y1,D).}$

\smallskip
\noindent
\textsc{Definition \& Folding.} 
Clauses 20 and 21 can be folded by using clause~17, thereby deriving:

\smallskip

\noindent
\makebox[6mm][r]{22.~}$\mathtt{new2(M,N,X,Y,Z,D)\,  \texttt{:-}\, M\!\geq\! 1,~~ N\!\geq\! 1,~~ X\!>\!Y,~~ X1\!=\!X\!-\!Y,~~ Z\!\neq\!D,~~ new2(M,N,X1,Y,Z).}$

\noindent
\makebox[6mm][r]{23.~}$\mathtt{new2(M,N,X,Y,Z,D)\,  \texttt{:-}\, M\!\geq\! 1,~~
N\!\geq\! 1,~~ X\!<\!Y,~~ Y1\!=\!Y\!-\!X,~~ Z\!\neq\!D,~~ new2(M,N,X,Y1,Z).}$

\smallskip
\noindent
No new predicate definition is introduced, and the \textit{Transform} strategy exits
the while-loop. The final program~$\textit{TransfP}$ is the set $\{18,19,22,23\}$ of clauses that contains no constrained facts. Hence both predicates $\mathtt{incorrect}$ and $\mathtt{new2}$ are useless and all clauses of~$\textit{TransfP}$ can be
deleted. Thus, the \textit{Transform} strategy
terminates with $\textit{TransfP}=\emptyset$ and we conclude that the imperative 
program \textit{gcd} is
correct with respect to the given properties
$\varphi_{\mathit{init}}$ and $\varphi_{\mathit{error}}$.

\section{Verifying Correctness of Array Programs}
\label{sec:array_properties}
In this section we apply our verification method of
Section~\ref{sec:verification_method}
for proving properties of a program, called {\it{arraymax}}, 
which computes the maximal element of an array.
Let us consider the following incorrectness triple
$\bbL\varphi_{\mathit{init}}(i,n,a,\textit{max})\bbR ~\mathit{arraymax}~
\bbL\varphi_{\mathit{error}}(n,a,\textit{max})\bbR$,
where:

\noindent
-- $\varphi_{\mathit{init}}(i,n,a,\textit{max})$ is~ 
$ i\!=\!0~ \andw ~n\!=\!{\mathit{dim}}(a)~  \andw ~n\!\geq\! 1~ \andw ~max\!=\!a[i]$,\nopagebreak

\noindent
\makebox[45mm][l]{-- $\mathit{arraymax}$ is the program} 

\noindent
\makebox[6mm][l]{}
$\ell_{0}\!:~\mathtt{while}~(i\!<\!n)~\{$\nopagebreak
$\mathtt{if}~(a[i] > max)~~max=a[i];$~~~
$i=i\!+\!1\,\};$\nopagebreak

\noindent
\makebox[6mm][l]{} 
$\ell_{h}\!:~\mathtt{halt}$

\noindent
-- $\varphi_{\mathit{error}}(n,a,\textit{max})$ is~
$\exists k ~(0\!\leq\! k\! <\! n \andw a[k]\!>\!\mathit{max})$.

\smallskip
\noindent
If this triple holds, then the value of
${\mathit{max}}$ computed by the program 
$\mathit{arraymax}$ is not the maximal element of the given
array $a$ with $n~(\geq 1)$ elements.

We start off by constructing a CLP program $T$ which encodes the incorrectness triple.
This construction is done as indicated in Section~\ref{sec:translating-correctness-into-CLP} and, in 
particular, the clauses for the predicates~${\mathtt{phiInit}}$ and ${\mathtt{phiError}}$
are generated from the formulas~$\varphi_{\mathit{init}}$ and~$\varphi_{\mathit{error}}$.

\smallskip
As indicated in Section~\ref{sec:verification_method}, the program \textit{arraymax} is 
translated into a set of CLP facts defining the predicate \texttt{at}.
The predicates $\mathtt{initConf}$  and $\mathtt{errorConf}$
specifying the initial and the error configurations, respectively, are defined by the 
following clauses:

\smallskip
\noindent
1.~$\mathtt{initConf(cf(cmd(0,asgn(int(x),int(0))),}$

\hspace{8mm}$\mathtt{[[int(i),I],[int(n),N],[array(a),(A,N)],[int(max),Max]]))}$ {\tt{:-}} 
$\mathtt{phiInit(I,N,A,Max).}$

\noindent
2.~$\mathtt{errorConf(cf(cmd(h,halt),}$

\hspace{8mm}$\mathtt{[[int(i),I],[int(n),N],[array(a),(A,N)],[int(max),Max]]))}$ {\tt{:-}} 
$\mathtt{phiError(N,A,Max).}$

\noindent
3.~$\mathtt{phiInit(I,N,A,Max)}$ {\tt{:-}} 
$\mathtt{I\!=\!0, ~~N\!\geq\!1, ~~read((A,N),I,Max)}.$

\noindent
4.~$\mathtt{phiError(N,A,Max)}$ {\tt{:-}}
$\mathtt{K\!\geq\! 0, ~~N\!>\!K, ~~Z\!>\!Max, ~~read((A,N),K,Z).}$

\smallskip
\noindent
Next, we apply Step~(A) of our verification method which consists in removing the interpreter
from program $T$. By applying the \textit{Transform} strategy as indicated
in Section~\ref{sec:verification_method}, we obtain the following program $\mathit{T_A}$:

\smallskip
\noindent
5.~$\texttt{incorrect}$ {\tt{:-}} $\mathtt{I\!=\!0, ~~N\!\geq\!1, ~~read((A,N),I,Max),~~ new1(I,N,A,Max).}$
  
\noindent
6.~$\mathtt{new1(I,N,A,Max)}$ {\tt{:-}}
$\mathtt{I1\!=\!I\!+\!1, ~~I\!<\!N, ~~I\geq 0, ~~M\!>\!Max,}$ 
$\mathtt{~read((A,N),I,M), ~~new1(I1,N,A,M).}$

\noindent
7.~$\mathtt{new1(I,N,A,Max)}$ {\tt{:-}}
$\mathtt{I1\!=\!I\!+\!1, ~~I\!<\!N, ~~I\!\geq\!0, ~~M\!\leq\!Max,}$ 
$\mathtt{~read((A,N),I,M), ~~new1(I1,N,A,Max).}$\nopagebreak

\noindent
8.~$\mathtt{new1(I,N,A,Max)}$ {\tt{:-}} $\mathtt{I\!\geq\! N,~~  
K\!\geq\! 0, ~~N\!>\!K, ~~Z\!>\!Max, ~~read((A,N),K,Z).}$

\smallskip
\noindent
We have that $\mathtt{new1(I,N,A,Max)}$ encodes the reachability 
of the error configuration from a configuration where the program
variables $\mathit{i, n, a, max}$ are bound to $\mathtt{I,N,A,Max}$,
respectively.

In order to propagate the error property, similarly to the example of
Section~\ref{sec:verification_method}, we first `reverse' program $\mathit{T_A}$ and we get the following program~$T_A^{\textit{rev}}$:

\smallskip
\noindent
rev1.~~\makebox[17mm][l]{$\mathtt{incorrect}$} {\tt{:-}} $\mathtt{b(U), ~~r2(U)}$.\nopagebreak

\noindent 
rev2.~~\makebox[8mm][l]{$\mathtt{r2(V)}$} {\tt{:-}} 
  $\mathtt{trans(U,V), ~~r2(U)}$.\nopagebreak  

\noindent
rev3.~~\makebox[8mm][l]{$\mathtt{r2(U)}$} {\tt{:-}} 
$\mathtt{a(U)}$.

\smallskip
\noindent
where predicates {\tt a(U)}, {\tt b(U)}, and {\tt trans(U,V)} are defined as follows:

\smallskip
\noindent
s4.~$\mathtt{a([new1,I,N,A,Max])}$ {\tt{:-}} 
$\mathtt{I\!=\!0,~~ N\!\geq\!1, ~~read((A,N),I,Max)}$

\noindent
s5.~$\mathtt{trans([new1,I,N,A,Max],[new1,I1,N,A,M])}$\, {\tt{:-}}\,

\hspace{40mm}$\mathtt{I1\!=\!I\!+\!1, ~~I\!<\!N, ~~I\!\geq\! 0, ~~M\!>\!Max, ~~read((A,N),I,M)}$.

\noindent
s6.~$\mathtt{trans([new1,I,N,A,Max],[new1,I1,N,A,Max])}$\, {\tt{:-}}\,

\hspace{40mm}$\mathtt{I1\!=\!I\!+\!1,~~ I\!<\!N,~~ I\!\geq\! 0,~~ M\!\leq\! Max, ~~read((A,N),I,M)}$.

\noindent
s7.~$\mathtt{b([new1,I,N,A,Max])}$ {\tt{:-}} 
$\mathtt{I\!\geq\! N, ~~K\!\geq\! 0, ~~K\!<\!N, ~~Z\!>\!Max, ~~read((A,N),K,Z)}$.

\smallskip
\noindent
For the application of Step (B) of the verification method
we assume that the predicates
\texttt{incorrect}, \texttt{b}, \texttt{r2},  \texttt{trans}, and \texttt{a} are 
high predicates, while the predicate \texttt{read} is a low predicate.

For the \textsc{Goal Replacement}
phase we use the following law, which is a consequence of the fact that an array
is a finite function:

\smallskip
\noindent
\makebox[16mm][l]{(Law~L1)~}
$\mathtt{read((A,N)\!,K,Z), ~~read((A,N),I,M)} ~\leftrightarrow$

\hspace{29mm}$(\mathtt{K\!=\!I,~~ Z\!=\!M,~~ read((A,N),K,Z)}) ~~~~\vee~~ $

\hspace{29mm}$(\mathtt{K\!\neq\! I,~~ read((A,N),K,Z),~~ read((A,N),I,M))}$

\smallskip
\noindent
In general, when applying the \textit{Transform} strategy in the case of array programs, some additional 
laws may be required (see, for instance,~\cite{Br&06}).
For the \textsc{Definition}~\&~\textsc{Folding} phase we use
a particular generalization operator, called
\mbox{\it{WidenSum}}~\cite{Fi&13a}, which is a variant of the classical widening operator~\cite{CoC77} and behaves as follows.

Given any atomic constraint $a$, let us denote by $\textit{sumcoeff}(a)$
the sum of the absolute values of the coefficients of $a$.
Given any two constraints~$c$ and~$d$, ${\mathit{WidenSum}}(c,d)$ returns 
the conjunction of: (i) all atomic constraints $a$ in $c$ such that $d\sqsubseteq a$, 
and (ii) all atomic constraints~$b$ in $d$ such that $\textit{sumcoeff}(b)\!\leq\! \textit{sumcoeff}(e)$ 
for some atomic constraint $e$ in $c$.

\smallskip
\noindent Let us now apply Step (B) of the verification method.

\smallskip
\noindent
\textsc{Unfolding}.  
We start off by unfolding clause~rev1 w.r.t.~the atom $\mathtt{b(U)}$, and we get:

\smallskip
\noindent
9.~$\mathtt{incorrect}$\,{\tt{:-}}\, 
$\mathtt{I\!\geq\!N,~~ K\!\geq\! 0,~~ K\!<\!N, ~~Z\!>\!Max,~~ read((A,N),K,Z),~~ r2([new1,I,N,A,Max])}$.

\smallskip
\noindent
The \textsc{Goal Replacement} and \textsc{Clause Removal} phases leave unchanged the set of clauses 
we have derived so far.

\smallskip
\noindent
\textsc{Definition \& Folding.} In order to fold clause~9 we introduce the 
following clause:

\smallskip
\noindent
10.~$\mathtt{new2(I,N,A,Max,K,Z)}$ {\tt{:-}} 
$\mathtt{I\!\geq\!N, ~~K\!\geq\! 0,~~ K\!<\!N,~~ Z\!>\!Max,~ read((A,N),K,Z),~~ r2([new1,I,N,A,Max])}$.

\smallskip
\noindent
By folding clause 9 using clause~10, we get:

\smallskip
\noindent
11. $\mathtt{incorrect}$ {\tt{:-}} 
$\mathtt{I\!\geq\!N, ~~K\!\geq\! 0, ~~K\!<\!N, ~~Z\!>\!Max, ~~new2(I,N,A,Max,K,Z)}$.

\smallskip
\noindent
Now we proceed by performing a second iteration of the body of the while-loop
of the \textit{Transform} strategy because clause~10 is in {\it{InDefs}}.

\smallskip

\noindent
\textsc{Unfolding}. 
By unfolding clause~10 w.r.t.~the atom $\mathtt{r2([new1,I,N,A,Max])}$,
we get the following clauses:

\smallskip
\noindent
12.~$\mathtt{new2(I,N,A,Max,K,Z)}$ {\tt{:-}} 
$\mathtt{I\!\geq\!N, ~~K\!\geq\! 0,~~ K\!<\!N,~~ Z\!>\!Max,~~}$

\hspace{22mm}$\mathtt{read((A,N),K,Z), ~~trans(U,[new1,I,N,A,Max]), ~~r2(U).}$

\noindent
13.~$\mathtt{new2(I,N,A,Max,K,Z)}$ {\tt{:-}} 
$\mathtt{I\!\geq\!N, ~~K\!\geq\! 0, ~~K\!<\!N,~~ Z\!>\!Max,~~ read((A,N),K,Z),~~ a([new1,I,N,A,Max])}$.

\smallskip
\noindent
By unfolding clause 12 w.r.t.~$\mathtt{trans(U,[new1,I,N,A,Max])}$,
 we get:

\smallskip
\noindent
14.~$\mathtt{new2(I1,N,A,M,K,Z)}$ {\tt{:-}}
$\mathtt{I1\!=\!I\!+\!1, ~~N\!=\!I1, ~~K\!\geq\! 0, ~~K\!<\!I1, ~~M\!>\!Max, ~~Z\!>\!M,}$ 

\hspace{22mm}$\mathtt{read((A,N),K,Z), ~~read((A,N),I,M), ~~r2([new1,I,N,A,Max]).}$

\smallskip
\noindent
15.~$\mathtt{new2(I1,N,A,Max,K,Z)}$ {\tt{:-}}
$\mathtt{I1\!=\!I\!+\!1,~~ N\!=\!I1,~~ K\!\geq\! 0,~~ K\!<\!I1,~~ M\!\leq\!Max,~~ Z\!>\!Max,}$\nopagebreak

\hspace{22mm}$\mathtt{read((A,N),K,Z), ~~read((A,N),I,M), ~~r2([new1,I,N,A,Max]).}$

\smallskip
\noindent
By unfolding clause 13 w.r.t.~$\mathtt{a([new1,I,N,A,Max])}$, 
we get an empty set of clauses (indeed, the
constraint $\mathtt{I\!\geq\!N,~ I\!=\!0,~ N\!\geq\!1}$ is  unsatisfiable).

\smallskip
\noindent
\textsc{Goal Replacement}. This phase performs the following two steps:
\noindent
(i)~it replaces the conjunction of atoms `$\mathtt{read((A,N),K,Z)},~\mathtt{read((A,N),I,M)}$'
occurring in the body of clause~14
by the right hand side of Law~L1, and then
(ii)~it splits the derived clause into the following two clauses, 
each of which corresponds to a disjunct of that right hand side.
 
\smallskip
\noindent
14.1~$\mathtt{new2(I1,N,A,M,K,Z)}$ {\tt{:-}}
$\mathtt{I1\!=\!I\!+\!1, ~~N\!=\!I1, ~~K\!\geq\! 0, ~~K\!<\!I1, ~~M\!>\!Max, ~~Z\!>\!M,}$

\hspace{22mm}$\mathtt{{K\!=\!I, ~~Z\!=\!M}}$,~~ $\mathtt{read((A,N),K,Z)}$,~~
$\mathtt{r2([new1,I,N,A,Max])}$.

\noindent
14.2~$\mathtt{new2(I1,N,A,M,K,Z)}$ {\tt{:-}}
$\mathtt{I1\!=\!I\!+\!1, ~~N\!=\!I1, ~~K\!\geq\! 0, ~~K\!<\!I1, ~~M\!>\!Max, ~~Z\!>\!M,}$ 

\hspace{22mm}$\mathtt{{K\!\neq\!I}}$,~~
$\mathtt{read((A,N),K,Z)}$,~~ $\mathtt{read((A,N),I,M)}$,~~ $\mathtt{r2([new1,I,N,A,Max])}$.

\smallskip
\noindent
\textsc{Clause Removal}. 
The constraint ~`$\mathtt{Z\!>\!M, ~Z\!=\!M}$'~ in the body of clause 14.1 
is unsatisfiable. Therefore, this clause is removed from \textit{TranfP}.  
From clause~14.2, by replacing `$\mathtt{K\!\not = \!I}$' by `$\mathtt{K\!<\!I ~\orv~ K\!>\!I}$' and simplifying the constraints, we get:

\smallskip
\noindent
16.~$\mathtt{new2(I1,N,A,M,K,Z)}$ {\tt{:-}}
$\mathtt{I1\!=\!I\!+\!1, ~~N\!=\!I1, ~~K\!\geq\! 0, ~~K\!<\!I, 
~~M\!>\!Max, ~~Z\!>\!M,}$

\hspace{22mm}$\mathtt{read((A,N),K,Z)}$,~~
$\mathtt{read((A,N),I,M)}$,~~
$\mathtt{r2([new1,I,N,A,Max])}$.

\smallskip
\noindent
By performing from clause~15 a sequence of goal replacement and clause removal transformations
similar to that we have performed from clause~14,
we get the following clause:

\smallskip
\noindent
17.~$\mathtt{new2(I1,N,A,Max,K,Z)}$ {\tt{:-}} 
$\mathtt{I1\!=\!I\!+\!1,~~ \,N\!=\!I1,~~ \,K\!\geq\! 0,~~ K\!<\!I,~~ M\!\leq\!Max,~~ Z\!>\!Max,}$ 

\hspace{22mm}$\mathtt{read((A,N),K,Z)}$,~~
$\mathtt{read((A,N),I,M)}$,~~
$\mathtt{r2([new1,I,N,A,Max])}$.

\smallskip
\noindent
\textsc{Definition \& Fold}. 
The comparison between the definition clause~10 we have 
introduced above, and clauses 16 and~17 
which we should fold, shows the risk of introducing
an unlimited number of definitions whose body contains the atoms 
$\mathtt{read((A,N),K,Z)}$ and $\mathtt{r2([new1,I,N,A,Max])}$.
Thus, in order to fold clauses 16 and 17,
we introduce the following new definition:

\smallskip
\noindent
18.~$\mathtt{new3(I,N,A,Max,K,Z)}$ {\tt{:-}}
$\mathtt{K\!\geq\! 0, ~K\!<\!N, ~K\!<\!I, ~Z\!>\!Max,~ read((A,N),K,Z)), ~r2([new1,I,N,A,Max])}$.

\smallskip
\noindent
The constraint in the body of this clause is obtained by generalizing:
(i) the projection of the constraint in the body of clause~16 on 
the variables  $\mathtt{I,N,A,Max,K,Z}$ (which are the variables 
of clause 16 that occur
in the atoms $\mathtt{read((A,N),K,Z)}$ and $\mathtt{r2([new1,I,N,A,Max])}$), and
(ii) the constraint occurring in the body of clause~10.
This generalization step can be seen as an application
of the above mentioned \mbox{\textit{WidenSum}} generalization operator.
The same definition clause 18 could also be derived by
generalizing the projection of the
constraint in the body of clause 16 (instead of 17) and
the constraint occurring in the body of clause~10. 

Thus, by folding clause 16 and clause 17 
using clause 18 we get:\nopagebreak

\smallskip
\noindent
19.~$\mathtt{new2(I1,N,A,Max,K,Z)}$ {\tt{:-}}
$\mathtt{I1\!=\!I\!+\!1, ~~N\!=\! I1, ~~K\!\geq\!0, ~~K\!<\!I, ~~M\!>\!Max, ~~Z\!>\!M,}$\nopagebreak 

\hspace{22mm}$\mathtt{read((A,N),I,M),\ new3(I,N,A,Max,K,Z)}$.

\noindent
20.~$\mathtt{new2(I1,N,A,M,K,Z)}$ {\tt{:-}} 
$\mathtt{I1\!=\!I\!+\!1, ~~N\!=\! I1, ~~K\!\geq\!0, ~~K\!<\!I,  ~~M\!\leq\!Max, ~~Z\!>\!Max,}$\nopagebreak 

\hspace{22mm}$\mathtt{read((A,N),I,M),\ new3(I,N,A,Max,K,Z)}$.

\smallskip
\noindent
Now we perform the third iteration of the body of the while-loop
of the strategy.

\smallskip
\noindent
\textsc{Unfolding}, \textsc{goal replacement}, and  \textsc{clause removal}.
By unfolding, goal replacement, and clause removal,
from clause 18 we get:

\smallskip
\noindent
21.~$\mathtt{new3(I1,N,A,M,K,Z)}$ {\tt{:-}} 
$\mathtt{I1\!=\!I\!+\!1, ~K\!\geq\!0, ~K\!\!<\!I, ~~N\!\geq\!I1, ~~M\!>\!Max, ~~Z\!>\!M,}$ 
          
\hspace{22mm}$\mathtt{read((A,N),K,Z),~~ read((A,N),I,M),~~ r2([new1,I,N,A,Max]).}$

\noindent
22.~$\mathtt{new3(I1,N,A,Max,K,Z)}$ {\tt{:-}} 
$\mathtt{I1\!=\!I\!+\!1, ~K\!\geq\!0, ~~K\!\!<\!I, ~~N\!\geq\!I1, ~~M\!\leq\!Max, ~~Z\!>\!Max,}$ 

\hspace{22mm}$\mathtt{read((A,N),K,Z), ~~read((A,N),I,M), ~~r2([new1,I,N,A,Max]).}$

\smallskip

\noindent
\textsc{Definition \& Folding}. In order to fold clauses 21 and 22,
we do not need to introduce any new definition.
Indeed, it is possible to fold these clauses by using 
clause 18, thereby obtaining:

\smallskip
\noindent
23.~$\mathtt{new3(I1,N,A,M,K,Z)}$ {\tt{:-}} 
$\mathtt{I1\!=\!I\!+\!1, ~~K\!\geq\!0,~~
K\!\!<\!I,~~ N\!\geq\!I1,~~ M\!>\!Max,~~ Z\!>\!M,}$
 
\hspace{22mm}$\mathtt{read((A,N),I,M),~~ new3(I,N,A,Max,K,Z)}$.

\noindent
24.~$\mathtt{new3(I1,N,A,Max,K,Z)}$ {\tt{:-}} 
$\mathtt{I1\!=\!I\!+\!1,~~ K\!\geq\!0,~~
K\!\!<\!I, ~~N\!\geq\!I1,~~ M\!\leq\!Max,~~ Z\!>\!Max,\ }$ 

\hspace{22mm}$\mathtt{read((A,N),I,M),~~ new3(I,N,A,Max,K,Z)}$.

\smallskip
\noindent
Since no clause to be processed is left (because $\textit{InDefs}\!=\!\emptyset$),
the \textit{Transform} strategy exits the  outermost while-loop,
and the program derived is the set $\{11, 19, 20, 23, 24\}$ of clauses.
No clause in this set is a constrained fact, and hence by  
\textsc{removal of useless clauses} we get the final program~$\mathit{T_B}$ 
consisting of the empty set of clauses.
Thus, $\mathit{arraymax}$ is correct with respect to the given properties
$\varphi_{\mathit{init}}$ and $\varphi_{\mathit{error}}$.


\section{Related Work and Conclusions}
\label{sec:conclusions}
The verification framework introduced in this paper is an extension 
of the framework presented in~\cite{De&13a}, where CLP and 
\textrm{iterated specialization} have been used to define a general 
verification framework which is parametric with respect to the 
programming language
and the logic used for specifying the properties of interest. 
The main novelties we have introduced in this paper are the following: 
(i)~we consider imperative programs acting also on array variables, and 
(ii)~we consider a more expressive specification language, 
which allows us to write properties involving elements
of arrays and, in general, fields of complex data
structures.

In order to deal with these additional features (i)~we have defined 
the operational semantics for array manipulation, and 
(ii)~we have considered powerful transformation rules,
such as conjunctive definition, conjunctive folding, and 
goal replacement.
The transformation rules and some  strategies for their application
have been implemented in 
the MAP transformation system~\cite{MAP}, so as to perform
proofs of program correctness in a semi-automated way.

Our approach has many connections to various
techniques developed in the field of static program analysis,
to which David Schmidt has given an outstanding contribution,
specially in clarifying its relationships with methodologies
like denotational semantics, abstract interpretation, and model checking
(see, for example,~\cite{Sch98}).

Now we briefly overview the 
verification techniques that are particularly relevant to 
the method described in the present paper, 
and make use of logic programming, constraints,
abstract interpretation, and automated theorem proving.

The use of logic programming techniques for program analysis is not novel. 
For instance, Datalog (the function-free
fragment of logic programming) has been proposed 
for performing various types of program analysis such as
{\em dataflow analysis}, {\em shape analysis}, and {\em points-to analysis}
\cite{BrS09,Rep98,WhL04}.
In order to encode the properties of interest into Datalog,
all these analysis techniques make an abstraction of the program semantics.
In contrast,  our transformational method manipulates a CLP program
which encodes the full semantics (up to a suitable level of detail)
of the program to be analyzed.
An advantage of our approach is that the output of a transformation-based
analysis is a new program which is {\em equivalent} to the initial
one, and thus the analysis can be iterated to the desired degree of
precision.

Constraint logic programming has been successfully 
applied to perform model checking of both finite~\cite{NiL00} and infinite 
state systems~\cite{DeP99,Fi&01a,Fi&13a}.
Indeed, CLP turns out to be suitable for expressing both 
(i)~the symbolic executions and (ii)~the invariants  
of imperative programs~\cite{Ja&11b}.
Moreover, there are powerful CLP-based tools, such as ARMC~\cite{PoR07}, 
TRACER~\cite{Ja&11c}, 
and HSF~\cite{HSF} that can be used for performing model checking of 
imperative programs.
These tools are fully automatic, but they are applicable
to classes of programs and properties that are much more limited
than those considered in this paper.
We have shown in~\cite{De&13a} that, by focusing on
verification tasks similar to those considered by ARMC, TRACER, and HSF,
we can design a fully automatic, transformation-based verification technique
whose effectiveness is competitive with respect to the one of 
the above mentioned tools.

The connection between imperative programs and 
constraint logic programming has been
investigated in~\cite{Fla04,Pe&98}. 
The verification method presented in~\cite{Fla04} is based on 
a semantics preserving translation from an imperative 
language with dynamic data structures and recursive 
functions into CLP. This translation reduces the verification of the (in)correctness
of imperative programs to a problem of constraint satisfiability within standard CLP systems.
A method based on specialization of CLP programs for performing the analysis 
of imperative programs has been presented in~\cite{Pe&98}. 
In this work the authors first present a CLP interpreter which defines
 the operational semantics of a simple imperative language. 
Then, given an imperative program, say $\textit{prog}$, they specialize that interpreter 
with respect to a CLP translation of~$\textit{prog}$ thereby getting a residual, specialized 
CLP program~$P_{\!\mathit{sp}}$. Finally, a static analyzer for CLP programs is applied 
to~$P_{\!\mathit{sp}}$ and its results are used to annotate 
the original imperative program with invariants.
In~\cite{HeG06} a very similar methodology 
is applied for the verification of low-level programs
for PIC microcontrollers.

The program specialization which is done during Step~(A) of our
 verification method
(that is, the removal of the interpreter) is very similar to the
specialization proposed in~\cite{Pe&98}.
However, having removed the interpreter, in order to verify the
correctness of the given imperative program, in Step (B) we apply again
program transformation and we not use static analyzers.

CLP approaches to the verification of program correctness have recently received 
a growing attention because of the development of 
very efficient \textrm{constraint solving} tools~\cite{Ryb10}.
These approaches include: (i)~the template-based invariant 
synthesis~\cite{Be&07}, and (ii)~the interpolant generation~\cite{Ry&10}. 
Related to this line of work, we would like to
mention the paper~\cite{Gr&12} where 
the authors propose a method for constructing verification 
tools starting from a given set of proof rules, called Horn-like
rules, specified by using constraints.

Our transformational method for verifying properties of array programs
is related to several methods based on abstract interpretation and predicate abstraction.
In~\cite{HaP08}, which builds upon~\cite{Go&05}, 
relational properties among array elements are discovered
by partitioning the arrays into symbolic slices 
and associating an abstract variable with each slice.
This approach offers a compromise between the efficiency of {\em array smashing}
(where the whole array is represented by a single abstract variable) 
and the precision of {\em array expansion}~\cite{Bl&02}
(where every array element is associated with a distinct abstract variable).
In~\cite{Co&11} the authors present a scalable, parameterized abstract 
interpretation framework for the automatic analysis of array programs based on slicing.
In~\cite{Gu&08} a powerful technique using 
template-based quantified abstract domains,  
is applied to successfully generate quantified invariants.
Other authors (see~\cite{FlQ02,LaB07}) use indexed predicate abstraction
for inferring universally quantified properties
about array elements.

Also theorem provers have been used for discovering invariants
in programs which manipulate arrays.
In particular, in~\cite{Br&06} a satisfiability decision procedure 
for a decidable fragment of a theory of arrays is presented.
That fragment is expressive enough to prove properties such
as sortedness of arrays.
In~\cite{JhM07,KoV09,McM08} 
the authors present some theorem provers which may generate 
array invariants and interpolants.
In~\cite{Se&09} a backward reachability analysis
based on predicate abstraction and the CEGAR technique  
is used for deriving invariants 
which are universally quantified formulas over array indexes, and
existing techniques for quantifier-free assertions
are adapted to the verification of array properties.

Finally, we would like to mention two more papers which deal with the problem
of verifying properties of array programs by using
Satisfiability Modulo Theory~(SMT) techniques.
Paper~\cite{La&13} presents  a constraint-based invariant 
generation method for generating universally quantified 
loop invariants over arrays, and 
paper~\cite{Al&12} proposes a model checker
for verifying universally quantified safety properties of arrays.

As future work, we plan to investigate the issue of designing a general,
fully automated strategy for guiding the application of the  
 program transformation rules we have considered in this paper.
In particular, we want to study the problem of devising unfolding strategies,
generalization operators, and goal-replacement techniques which are tailored to the
specific task of verifying program correctness. 
We also intend to extend the implementation described in~\cite{De&13a}
and to
perform more experiments for validating our
transformation-based verification approach by using some benchmark suites which 
include array programs. 

As a further line of future research, we plan to enrich our framework by
considering some more 
theories, besides the theory of the arrays, so that one can prove properties of 
programs acting on dynamic data structures such as lists and heaps.

\section{Acknowledgments}
We thank the anonymous referees for their constructive comments.
We would like to thank Anindya Banerjee, Olivier Danvy,
Kyung-Goo Doh, and
John Hatcliff for their kind invitation to contribute to
this symposium in honor of Dave Schmidt. 
Alberto recalls with great joy and gratitude
the time together with Dave in Edinburgh and in 
Rome.

\end{document}